\documentclass[a4paper, 12pt]{article}

\usepackage{styleBuding}
\usepackage{bm}
\usepackage{color,listings}
\usepackage[usenames,dvipsnames,svgnames,table]{xcolor}
\usepackage{amsthm, amsmath}
\usepackage{amssymb}
\usepackage{mathrsfs}
\usepackage{graphicx}
\usepackage{slashed}
\usepackage{soul}
\usepackage{multirow}
\usepackage{subfigure}
\usepackage{diagbox}
\usepackage{siunitx} % for physics units
\usepackage{url} %for link in @article
\usepackage[utf8]{inputenc}
\usepackage[figuresright]{rotating}
\definecolor{back}{HTML}{F8F8F8}
\usepackage{epstopdf}
\usepackage{lipsum}

\usepackage{booktabs}
\usepackage{array}

\title{Connecting Electroweak Fine-Tuning, Tens-of-GeV Dark Matter, and Exotic Higgs Decays in the GNMSSM}
\author{Zhaoxia Heng $^1$, Fei Li $^2$, Haijing Zhou $^1$}
\affiliation{$^1$ School of Physics, Henan Normal University, Xinxiang 453007, China\\
$^2$ School of Physics, Zhengzhou University, Zhengzhou 450000, China}
\emailAdd{zxheng@htu.edu.cn}
\emailAdd{hnufeili@163.com}
\emailAdd{zhouhaijing0622@163.com}

\abstract{
In the general next-to-minimal supersymmetric standard model (GNMSSM), we connect electroweak fine-tuning, a singlino dark matter candidate with mass around tens of GeV, and Exotic Higgs Decays, identifying a natural region characterized by the mass hierarchy $m_{\tilde{S}}<m_{\tilde{B}}<m_{\tilde{H}}$ and $\mu_{\rm tot}\in [200, 500]$ GeV. For fine-tuning measure $\Delta \leq 40$, the branching ratio $\mathrm{Br}(h\to a_sa_s)$ falls in the range 7\%$-$25\%, with the pseudoscalar mass $m_{a_s}$ lying mostly between 10 and 20 GeV.  This region splits into two viable scenarios consistent with LZ 2024 direct detection: one where the lightest CP-even Higgs $h_1$ is SM-like ($h_1$ case), and another where the second-lightest $h_2$ is SM-like ($h_2$ case). LHC searches for $h\to a_sa_s \to b\bar{b}\tau^+\tau^-$ exclude 12\% (10\%) of samples with 12 GeV $\leq m_{a_s} \leq$ 15.5 GeV in the $h_1$ ($h_2$) case, while a substantial fraction remains accessible at the HL-LHC in the $h_2$ case. Our analysis unifies naturalness, dark matter, and observable exotic Higgs decays, yielding concrete targets for future experiments.}

\begin{document}
\maketitle

\section{Introduction}
The nature of dark matter (DM) and the origin of the electroweak (EW) scale rank among the most profound open questions in particle physics \cite{Hambye:2007vf}. Weakly Interacting Massive Particles (WIMPs) remain among the most prominent DM candidates, owing to their ability to naturally reproduce the observed cosmic relic abundance \cite{Planck:2018vyg} through the thermal freeze-out mechanism.
Within the broad WIMP mass window, the tens-of-GeV regime carries special significance. It lies near the EW scale, thereby linking DM to the EW symmetry-breaking mechanism \cite{Arcadi:2017kky,Arcadi:2024ukq}, and it coincides roughly with the mass range in which direct detection experiments achieve their best sensitivity to the spin-independent (SI) cross section — e.g., LUX-ZEPLIN (LZ) reaches its minimum SI exclusion near 40 GeV \cite{LZ:2024zvo}. The region is thus a critical testing ground for DM models. 
However, recent experimental results have posed serious challenges to the WIMP parameter space in this mass region. The LZ 2024 data pushed the upper limits on SI and spin-dependent (SD) WIMP–nucleon cross sections down to the  $10^{-48}~\mathrm{cm}^2$ and $10^{-43}~\mathrm{cm}^2$ levels, respectively \cite{LZ:2024zvo}. This severely constrains many classic WIMP realizations, such as the minimal supersymmetric Standard Model (MSSM) \cite{Nilles:1983ge,Haber:1984rc,Gunion:1984yn,Martin:1997ns}. One generic way to relax such bounds is the ``secluded" DM picture \cite{Pospelov:2007mp}, in which DM is largely decoupled from the Standard Model (SM) and can maintain a successful thermal freeze-out while evading direct detection. Nevertheless, many well-motivated SUSY models struggle to accommodate a viable DM candidate without developing a severe fine-tuning problem once all current constraints are imposed. A systematic exploration of models that can predict light DM without excessive fine-tuning is therefore of paramount importance.

Within low-energy SUSY, the MSSM can provide a tens-of-GeV Bino-like lightest neutralino as WIMP DM candidate.
To evade the stringent LZ 2024 SD bounds, MSSM requires the Higgsino mass parameter $\mu \gtrsim$ 700 GeV for $\tan\beta \gtrsim 4$ \cite{Li:2025qkg}. 
Such a large $\mu$ strongly suppresses the couplings of DM to the $Z$ boson and Higgs bosons. To nevertheless reproduce the observed relic density, the DM mass must satisfy a resonance condition, $|m_{\tilde{\chi}_1^0}|\simeq m_Z/2$ or $m_h/2$, which introduces an additional fine‑tuning issue \cite{Baer:2013gva,Cao:2018rix,LI2024116591, Abdallah:2020yag}.
If $\mu$ grows too large, even resonant annihilation may fail to generate the observed abundance, potentially rendering the light-DM scenario infeasible. 
In this region ($\mu \gtrsim$ 700 GeV), fine-tuning $\Delta ^ {-1}$$\sim 0.8\%$   in reproducing the $Z$-boson mass is introduced~\cite{Baer:2013gva, Li:2025qkg}. 

The $Z_3$-symmetric next-to-MSSM ($Z_3$-NMSSM)~\cite{Miller:2003ay,Ellwanger:2009dp,Maniatis:2009re} extends the MSSM with a singlet superfield $\hat{S}$, which enriches the phenomenology and offers a dynamical solution to the $\mu$-problem in the MSSM. A tens-of-GeV DM candidate in the $Z_3$-NMSSM can be either bino- or singlino-dominated, with the singlino mass $m_{\tilde{S}}$ tied to the higgsino mass $\mu_{\rm eff}$ via the ratio $2\kappa/\lambda$.  
For bino-dominated DM, the constraint from LZ 2024 on the parameter space is the same as that in the MSSM. 
In this context, the most general SI and SD blind spot conditions for bino-like DM have been derived, for the first time, in the $Z_3$-NMSSM by considering the complete bino--higgsino--singlino neutralino system~\cite{Abdallah:2020yag}.
For a viable singlino-dominated DM below 100 GeV, one finds $\kappa \lesssim \lambda/4$ for $\mu_{\rm eff} \gtrsim 200~\text{GeV}$~\cite{Li:2025qkg}; 
dedicated blind spot conditions for singlino-dominated DM, including the case where the higgsino and singlino mass parameters carry opposite relative signs, have been analyzed in Ref.~\cite{Roy:2024yoh}.
Various LHC search strategies have also been proposed to probe such blind spot 
regions, e.g.\ hunting ewinos and a light singlet-like scalar in top squark cascade 
decays~\cite{Datta:2022bvg}, searches for radiative neutralino decays into soft 
photons~\cite{Roy:2024yoh,Arganda:2025fhx}, and machine-learning-enhanced 
analyses~\cite{Arganda:2025fhx}, including in the context of a strong first-order 
electroweak phase transition in the $Z_3$-NMSSM~\cite{Chatterjee:2022pxf}.
Furthermore, all DM couplings in the $Z_3$-NMSSM are strongly suppressed by either 
$\lambda$ or $\kappa$ to evade the stringent LZ 2024 bounds, and achieving the 
observed relic density requires severe cancellations or extreme resonance tuning. 
As a result, the $Z_3$-NMSSM is also highly fine-tuned in this mass regime.

A promising way around this obstacle is  offered by the general NMSSM (GNMSSM) \cite{Choi:2019yrv,Cao:2022ovk,Cao:2024axg,Meng:2024lmi,Cao:2021ljw,95GeV2023,Cao:2022chy}.
In contrast to $Z_3$-NMSSM, the GNMSSM explicitly breaks the discrete $Z_3$ symmetry  through the introduction of bilinear mass parameters $\mu$, $\mu\prime$ and the linear tadpole coefficient $\xi$.
This extension effectively disentangles the Singlino mass parameter $m_{\tilde{S}}$ from the Higgsino mass parameter $\mu_{\rm tot}$.
As a result, the coupling coefficient $\kappa$, which governs the DM self-annihilation in the secluded sector, can be chosen independently of $\lambda$, which controls the DM-nucleon scattering. 
This decoupling enables the GNMSSM to satisfy the observed relic density through channels such as $\chi_1^0\chi_1^0\to h_s a_s$, $h_sh_s$, $a_sa_s$ ($h_s$ and $a_s$ denote the singlet-dominated CP-even and CP-odd Higgs bosons) while naturally suppressing the SI scattering cross section to meet the LZ 2024 limits\cite{Cao:2024axg,Meng:2024lmi,Li:2025qkg}. 
Furthermore, a light $a_s$ with a mass ranging from several to tens of GeV strongly enhances the exotic decay $h \to a_s a_s$ \cite{Heng:2025wng,Li:2025qkg}, and such a light pseudoscalar has emerged as a leading candidate in the search for new physics at the LHC.
ATLAS and CMS have conducted multiple systematic searches covering typical final states such as $4b$~\cite{CMS:2024zfv}, $\mu\mu bb$~\cite{CMS2024,CMS:2017dmg,ATLAS:2018emt}, $\tau\tau bb$~\cite{CMS:2018zvv,ATLAS2024}, $\gamma\gamma\tau\tau$~\cite{ATLAS:2024nnm}, and $4\mu$~\cite{ATLAS:2015hpr,CMS:2012qms,CMS:2015nay,ATLAS:2018coo,CMS:2018jid}.
Beyond these resolved multi-object final states, $h\to a_sa_s$ additionally gives, for $m_{a_s}\sim 10$--20 GeV, a merged wide jet topology in which each $a_s$ appears as a single wide jet with a two-prong substructure \cite{Dobrescu:2026fok}.
Thus, $h \to a_s a_s$ not only bridges naturalness, WIMP-like light dark matter, and Higgs physics, but also offers a theoretically well-motivated channel with clear and collider-accessible signatures.

In this work, we aim to fill the gap of Ref.\cite{Li:2025qkg} and \cite{Heng:2025wng} by establishing a direct connection between the degree of fine-tuning in electroweak symmetry breaking, the viability of tens-of-GeV singlino-dominated DM, and the exotic decay branching ratio of the SM-like Higgs boson into a pair of light CP-odd Higgs singlets.
It is worth emphasizing that the viable parameter samples adopted in this work are drawn from the same scan as the previous analysis \cite{Li:2025qkg}, which focused on natural realization of tens-of-GeV DM in the GNMSSM.
And the current work performed a completely new post-processing analysis on these samples. Specifically, we computed the fine-tuning measurement $\Delta$ and the branching ratio of the exotic Higgs decay $\mathrm{Br}(h\to a_s a_s)$, which were not studied in \cite{Li:2025qkg}. Therefore, while the input data set overlaps, all numerical results, correlation maps, and phenomenological conclusions presented in Section 3 are entirely new.
For the first time, we quantitatively map $\Delta$ against the branching ratio $\mathrm{Br}(h\to a_s a_s)$ and connect the degree of naturalness to a potentially observable collider signature.

The remainder of this paper is organized as follows.
Sec.2 introduces the notations in the GNMSSM framework and provides analytical expressions relevant to DM-nucleon SI scattering cross sections. We also describe the calculation of fine-tuning in electroweak symmetry breaking. 
Sec.3 presents the numerical results concerning naturalness, DM and exotic Higgs decays.  Finally, Sec.4 summarizes the main conclusions of this study.

\section{Theoretical preliminaries}
\subsection{The GNMSSM and notations}
The GNMSSM extends the MSSM by a gauge-singlet superfield $\hat{S}$.
The most general renormalizable superpotential and corresponding soft-breaking terms are given by \cite{Ellwanger:2009dp, Li:2025qkg}:
\begin{eqnarray}
W_{\rm GNMSSM} &=& W_{\text{Yukawa} }+ \lambda \hat{S}\hat{H_u} \cdot \hat{H_d} + \frac{\kappa}{3}\hat{S}^3 + \mu \hat{H_u} \cdot \hat{H_d} + \frac{1}{2} \mu^{\prime} \hat{S}^2 + \xi\hat{S}, \\
-\mathcal{L}_{\rm soft} &= &\Bigg[\lambda A_{\lambda} S H_u \cdot H_d + \frac{1}{3} \kappa A_{\kappa} S^3+ m_3^2 H_u\cdot H_d + \frac{1}{2} {m_S^{\prime}}^2 S^2 + \xi_S S + h.c.\Bigg] \nonumber \\
   && + m^2_{H_u}|H_u|^2 + m^2_{H_d}|H_d|^2 + m^2_{S}|S|^2, 
\end{eqnarray}
where $W_{\rm Yukawa}$ includes the quark and lepton Yukawa interactions that are the same as those of the MSSM.
The dimensionless couplings $\lambda$ and $\kappa$ govern the Higgs doublet-singlet interaction and singlet cubic self-coupling, respectively. 
By introducing explicit $Z_3$-symmetry-breaking parameters $\mu$, $\mu^\prime$ and $\xi$, the inherent tadpole problem \cite{Ellwanger:1983mg, Ellwanger:2009dp} and cosmological domain-wall problem \cite{Abel:1996cr, Kolda:1998rm, Panagiotakopoulos:1998yw} in the $Z_3$-NMSSM can be resolved.
The parameters $\mu$ and $\mu^\prime$ can be naturally generated from spontaneous breaking of discrete $R$-symmetries ($\mathbb{Z}^R_4$ or $\mathbb{Z}^R_8$) at high scales \cite{Abel:1996cr, Lee:2010gv, Lee:2011dya, Ross:2011xv, Ross:2012nr}.
And the parameter $\xi$ can be absorbed by a redefinition of the Higgs singlet field $\hat{S}$ \cite{Ross:2011xv}. So we set $\xi = 0$ without loss of generality throughout this work.
These explicit $Z_3$-breaking terms not only alter the neutral Higgs mass spectrum, but also result in significantly richer phenomenology than those in the $Z_3$-NMSSM and MSSM.

After electroweak symmetry breaking, the neutral Higgs sector comprises three CP-even states $h_i$ ($i=1,2,3$) and two CP-odd states $a_j$ ($j=1,2$), while the neutralino sector contains five Majorana states $\tilde{\chi}_i^0 (i=1,2,...,5)$.
Following the conventions of our previous work \cite{Li:2025qkg}, we adopt the same Higgs and neutralino mass matrices and their diagonalization procedures, and thus do not reproduce them here.
We identify $h_s$ and $a_s$ as the singlet-dominated scalar and pseudoscalar (with components $|V^S|^2>0.5$), and $h$ as the SM-like Higgs boson.
The $h_1$ scenario corresponds to $h \equiv h_1$ with $m_{h_s} > m_h$, while the $h_2$ scenario corresponds to $h \equiv h_2$ with $m_{h} > m_{h_s}$.

The higgsino mass parameter $\mu_{\rm tot}$ and the singlino mass parameter $m_{\tilde{S}}$ play central roles in our analysis, and they are defined as 
\begin{eqnarray}
    \mu_{\rm tot}\equiv \lambda s +\mu,~~~~m_{\tilde{S}}\equiv 2\kappa s+\mu^\prime
\end{eqnarray}
where $s\equiv \langle S \rangle$ is the vacuum expectation value (VEV) of the singlet scalar $S$.
Crucially, unlike in the $Z_3$-NMSSM,  $\mu_{\rm tot}$ and $m_{\tilde{S}}$ are independent parameters in the GNMSSM.
This decoupling allows $\lambda$ to control the DM direct detection cross section while $\kappa$ governs the DM self-annihilation in the secluded sector, which is a key feature that distinguishes GNMSSM from $Z_3$-NMSSM and enables a natural tens-of-GeV lightest neutralino $\tilde{\chi}_1^0$ as WIMP DM candidate \cite{Li:2025qkg}.

According to \cite{Li:2025qkg}, the GNMSSM features two characteristic neutralino mass hierarchies: (i) $m_{\tilde{S}} < m_{\tilde{B}} < m_{\tilde{H}}$ with $M_1 = 150~\text{GeV}$, where the bino serves as the next-to-lightest supersymmetric particle (NLSP); and (ii) $m_{\tilde{S}} < m_{\tilde{H}} < m_{\tilde{B}}$ with $M_1 = 2~\text{TeV}$, where the higgsino is the NLSP. As demonstrated in \cite{Li:2025qkg}, these hierarchies impose markedly distinct LHC constraints on the total higgsino mass parameter $\mu_{\rm tot}$. In case (i), $\mu_{\rm tot}$ can be as low as $\sim 200~\text{GeV}$, preserving electroweak naturalness; in case (ii), $\mu_{\rm tot} \gtrsim 900~\text{GeV}$ is required, implying substantial fine-tuning in electroweak symmetry breaking. Extending the analysis of \cite{Li:2025qkg}, this work carries out a fully independent post-processing study of the corresponding parameter samples for both hierarchy regimes. Specifically, we compute the electroweak fine-tuning measure $\Delta$ and the branching ratios for exotic Higgs decays—quantities not evaluated in \cite{Li:2025qkg}.

\subsection{Dark Matter direct detection}
In the parameter space of interest, since the doublet-like Higgs boson $H$ is decoupled ($m_H\gg m_h$), the SI scattering cross section is dominated by the t-channel exchanges of the SM-like Higgs $h$ and the singlet-dominated scalar $h_s$.
In the limit of small $\lambda$, the SI cross section can be written as \cite{Meng:2024lmi}
\begin{eqnarray}
   \sigma^{\rm SI}_{\tilde{\chi}_1^0-N} \simeq 5\times 10^{-45}{\rm cm}^2(\frac{\mathcal{A}_h+\mathcal{A}_{h_s}}{0.1})^2 
\end{eqnarray}
where
\begin{eqnarray}
    \mathcal{A}_h &=& \left(\frac{125 \rm GeV}{m_h}\right )^2 V_h^{\rm SM}(V_h^{\rm SM}C_{\tilde{\chi}_1^0 \tilde{\chi}_1^0 H_{SM}}+V_h^{\rm S}C_{\tilde{\chi}_1^0 \tilde{\chi}_1^0 Re[S]}) \nonumber \\
    \mathcal{A}_{h_s} &=& \left(\frac{125 \rm GeV}{m_{h_s}}\right )^2 V_{h_s}^{\rm SM}(V_{h_s}^{\rm SM}C_{\tilde{\chi}_1^0 \tilde{\chi}_1^0 H_{SM}}+V_{h_s}^{\rm S}C_{\tilde{\chi}_1^0 \tilde{\chi}_1^0 Re[S]}) \label{amplitude}
\end{eqnarray}
When wino decouples, the couplings $C_{\tilde{\chi}_1^0 \tilde{\chi}_1^0 H_{SM}}$ and $C_{\tilde{\chi}_1^0 \tilde{\chi}_1^0 Re[S]}$ reduce to the simplified expression as follows,
\begin{eqnarray}
  C_{\tilde{\chi}_1^0 \tilde{\chi}_1^0 H_{SM}}&=& \sqrt{2}\lambda (N_{13}\sin\beta + N_{14}\cos\beta)N_{15} + g_1 (N_{13}\cos\beta - N_{14}\sin\beta)N_{11} \nonumber\\
  C_{\tilde{\chi}_1^0 \tilde{\chi}_1^0 Re[S]}&=& \sqrt{2}(\lambda N_{13}N_{14} - \kappa N_{15}N_{15}) \label{coupling-chih}
\end{eqnarray}
In the above equations, $|V_{h_i}^{\rm SM}|^2$ and $|V_{h_i}^{\rm S}|^2$ denote the fractions of the SM and singlet components in  $h_i$ ($i$= $h$ or $h_s$), respectively.
$N_{11}^2, N_{13}^2+ N_{14}^2,  N_{15}^2$ measure the bino, higgsino and singlino fractions in $\chi_1^0$, respectively. 

\subsection {Electroweak fine-tuning in the GNMSSM}\label{EW fine-tuning}
In the GNMSSM, the mass of $Z$ boson can be expressed as \cite{Baer:2012cf, Baer:2012uy}
\begin{eqnarray} \label{fine-tuning}
    m_Z^2 = 2\cdot\frac{m_{H_d}^2+\Sigma_d -(m_{H_u}^2+\Sigma_u)\tan^2\beta}{\tan^2\beta-1}- 2\mu_{\rm tot}^2  
\end{eqnarray}
where $\Sigma_d$ and $\Sigma_u$ are radiative corrections to the SUSY-breaking masses $m_{H_d}^2$ and $m_{H_u}^2$, respectively.
These corrections arise from the Higgs effective potential at loop level, with the dominant contributions coming from the Yukawa interactions of the third generation squarks \cite{Baer:2012cf}:
\begin{eqnarray}
    \Sigma_u \sim  \sum_{i=1}^2 \frac{3Y_t^2}{16\pi^2} m_{\tilde{t}_i}^2 (\ln \frac{m_{\tilde{t}_i}^2}{Q^2} -1), 
 ~~~~\Sigma_d \sim  \sum_{i=1}^2 \frac{3Y_b^2}{16\pi^2} m_{\tilde{b}_i}^2 (\ln \frac{m_{\tilde{b}_i}^2}{Q^2} -1) 
\end{eqnarray}
where $Q$ denotes the renormalization scale, and its optimized value is usually taken as $Q=\sqrt{ m_{\tilde{t}_1} m_{\tilde{t}_2}}$ with $ m_{\tilde{t}_1}$ and $m_{\tilde{t}_2}$ being the masses of the lighter and heavier top squarks, respectively.

A theory is deemed natural if each individual term on the right-hand side of Eq.(\ref{fine-tuning}) is of comparable magnitude to $m_Z^2$, thereby avoiding the need for large cancellations. To quantify the degree of naturalness, we adopt the standard Barbieri–Giudice criterion and define the fine-tuning measure as \cite{Barbieri:1987fn, Ellis:1986yg, Ross:2011xv, Ellwanger:2011mu, Cao:2018rix, Fowlie:2024nhs, vanBeekveld:2019tqp, Baer:2013gva, Baer:2026idw}
\begin{eqnarray} \label{delta}
    \Delta\equiv \max \{\Delta p_i\},~~~~~
    \Delta p_i\equiv\left|\frac{ \partial (\ln m_Z^2)}{\partial (\ln p_i^{\rm SUSY})}\right| 
\end{eqnarray}
where the relevant SUSY-scale parameters are
\begin{eqnarray}\label{psusy}
    p_i^{\rm SUSY} = m_{H_u}^2,  m_{H_d}^2, m_S^2, A_\lambda, A_\kappa, \xi_S, \mu, \mu^\prime, m_S^{\prime 2}, m_3^2, \lambda, \kappa, Y_t.
\end{eqnarray}

To compute the logarithmic derivatives in Eq.(\ref{delta}), we employ the three minimization equations of the scalar potential. With the dominant radiative corrections included, these equations take the form \cite{Ellwanger:2009dp,  Ross:2011xv}
\begin{eqnarray}\label{minimization equation}
E_1:  &\quad & m^2_{H_u} + \mu^2_{\text{tot}} + \lambda^2 v_d^2 + \frac{g_1^2+g_2^2}{4}(v_u^2-v_d^2) - \frac{v_d}{v_u}[\lambda s(A_\lambda+\kappa s)+m_3^2+\lambda \mu^\prime s]\nonumber\\
 && + \frac{3Y_t^4 v_u^2}{8\pi^2}\ln(m_{\tilde{t}_1}^2m_{\tilde{t}_2}^2/m_{\text{t}}^2) = 0, \nonumber\\
E_2: &\quad & m^2_{H_d} + \mu^2_{\text{tot}} + \lambda^2 v_u^2 + \frac{g_1^2+g_2^2}{4}(v_d^2-v_u^2) - \frac{v_u}{v_d}[\lambda s (A_\lambda+\kappa s)+m_3^2+\lambda\mu^\prime s] = 0, \nonumber\\
E_3: &\quad & m^2_{S} + m_S^{\prime 2} +\mu^{\prime 2} + \kappa A_\kappa s + 2\kappa^2 s^2 + 3\kappa s \mu^{\prime}+ \lambda^2(v_u^2+v_d^2) - 2\lambda\kappa v_u v_d \nonumber\\
&&+ \frac{\xi_S}{s}- \lambda \frac{v_u v_d}{s}(A_\lambda + \mu^{\prime})= 0,
\end{eqnarray}
where $g_1$, $g_2$ and the vevs $v_u$, $v_d$, $s$ can be expressed in terms of $m_Z$, $\tan\beta$ and $\mu_{\rm tot}$. 

For a given parameter set, the variations $\delta m_Z$ induced by shifts in $p_{i}^ {\mathrm {SUSY}}$ are obtained by solving the linearized system
\begin{eqnarray}
0 = \delta E _ {j} = \sum_ {i} \frac {\partial E _ {j}}{\partial p_{i} ^ {\mathrm {Susy}}} \delta p _ {i} ^ {\mathrm {Susy}} + \frac {\partial E _ {j}}{\partial M _ {Z}} \delta M _ {Z} + \frac {\partial E _ {j}}{\partial \tan \beta} \delta \tan \beta + \frac {\partial E _ {j}}{\partial \mu_ {\mathrm {eff}}} \delta \mu_ {\mathrm {tot}}   \label{deltaEj}
\end{eqnarray}
for $j=1,2,3$. Since all partial derivatives in Eq.(\ref{deltaEj}) can be evaluated analytically from Eq.(\ref{minimization equation}), the system yields a direct solution for $\delta m_Z$ in terms of $\delta p_{i}^ {\mathrm {SUSY}}$, thus allowing us to determine $\Delta$ according to Eq.(\ref{delta}).

It is instructive to identify the dominant source of fine-tuning in the GNMSSM. 
Eq.(\ref{fine-tuning}) shows that the mass of $Z$ boson receives a tree-level contribution proportional to $-\mu_{\rm tot}^2$.
For fixed soft masses, this implies $\partial m_Z^2/\partial \mu_{\rm tot}^2\simeq -2$, which naturally explains why the fine-tuning measurement $\Delta$ increases monotonically with $\mu_{\rm tot}$.
The quantity $\Delta^{-1}$ provides an intuitive measurement of the degree of tuning and $\Delta=$ 100 corresponds to 1\% fine-tuning.

\section{Numerical results}
The parameter samples analyzed in this study are drawn from the comprehensive parameter scan of the GNMSSM conducted in Ref.\cite{Li:2025qkg}.  That work employs the MultiNest algorithm ($n_{\rm live} = 8000$) to explore the parameter space, 
guided by a combined likelihood function that incorporates theoretical and experimental constraints, including the Planck 2018 relic dark matter density bound, 
the LZ 2024 direct detection upper limit, Higgs signal strength constraints implemented via \textsf{HiggsTools-1.2} \cite{Bahl:2022igd}, and B-physics observables. 
All samples have been rigorously tested against the latest ATLAS and CMS searches for electroweak supersymmetric particles using \texttt{SModelS-3.0.0}~\cite{Khosa:2020zar},
retaining only those points satisfying all LHC constraints. The complete scanning strategy, including prior settings, likelihood construction, and the full list of applied LHC analyses, has been detailed in \cite{Li:2025qkg} and will not be repeated here. 
This work directly uses these LHC-verified surviving points as input data to calculate the electroweak fine-tuning measure $\Delta$
\footnote{The value of $\Delta$ is computed using the modified version of the ftpar.f file from \texttt{NMSSMTools-6.2.2}~\cite{Ellwanger:2004xm, Ellwanger:2005dv}, as described in Section 2.2 of this paper.}, and performs new post-processing analyses. 
Specifically, we analyzed the parameter space features that simultaneously accommodate a natural electroweak scale, a viable TeV-scale singlino-dominated dark matter candidate, and a potentially observable signature of exotic Higgs decay.

\begin{figure}
\centering
\includegraphics[width=0.33\linewidth]{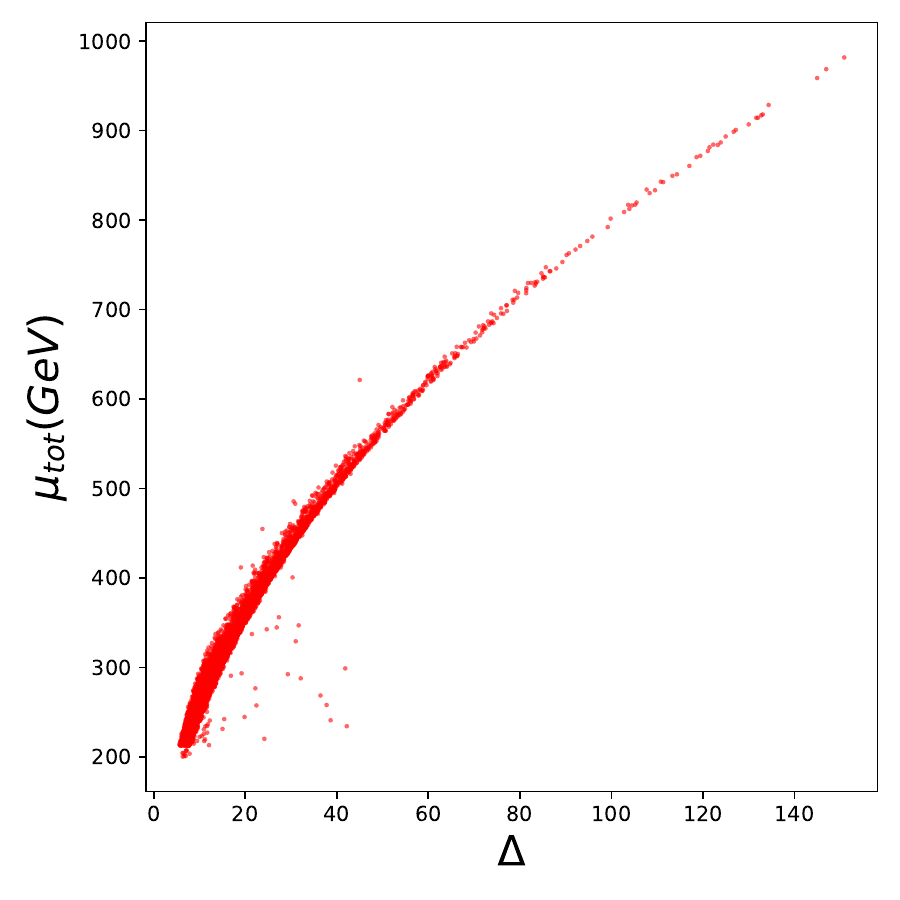}
   \hspace{-0.3cm}
\includegraphics[width=0.33\linewidth]{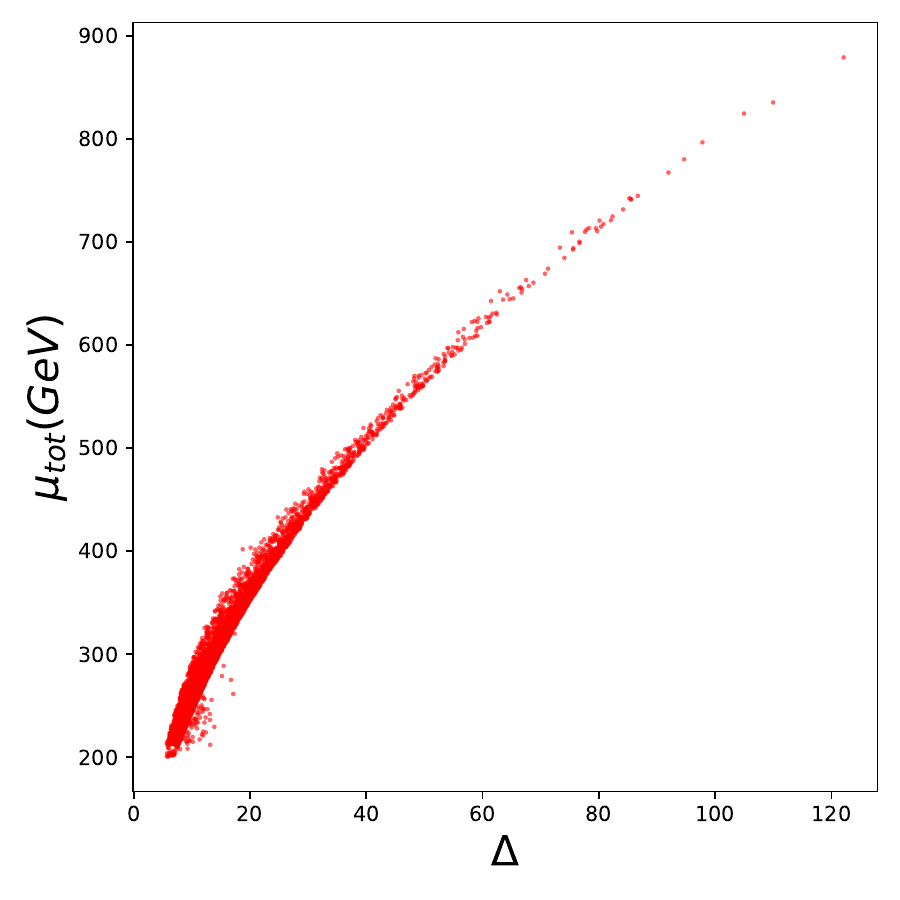}
   \hspace{-0.3cm}
\includegraphics[width=0.33\linewidth]{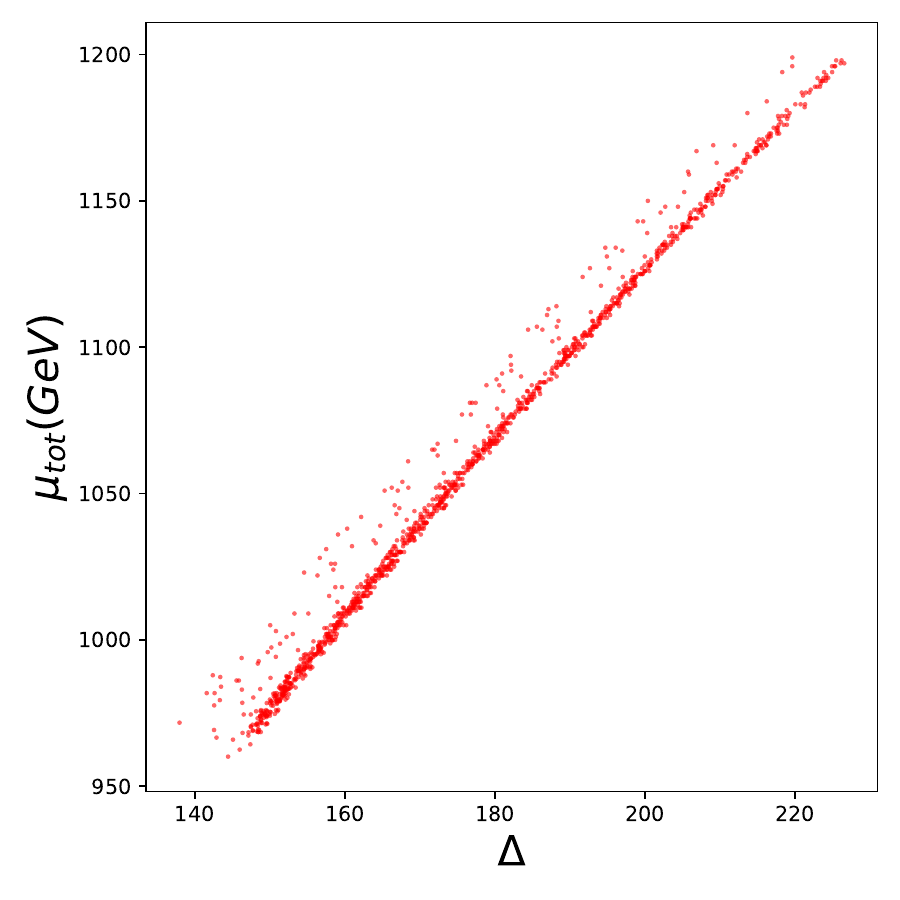}
\caption{Electroweak fine-tuning measure $\Delta$ vs. $\mu_{\text{tot}}$ for three scenarios: $h_1$-scenario with $M_1 = 150~\text{GeV}$ (left panel), $h_2$-scenario with $M_1 = 150~\text{GeV}$ (middle panel), and $h_1$-scenario with $M_1 = 2~\text{TeV}$ (right panel).}\label{fig1}
\end{figure}

Figure~\ref{fig1} presents the distribution of the fine tuning measure $\Delta$ as a function of $\mu_{\rm tot}$ for the surviving samples.
The left and middle panels correspond to the mass hierarchy $m_{\tilde{S}} < m_{\tilde{B}} < m_{\tilde{H}}$ with $M_1$= 150 GeV for the $h_1$ and $h_2$ scenarios, respectively.  
The right panel shows $m_{\tilde{S}} < m_{\tilde{H}} < m_{\tilde{B}}$ with $M_1$= 2 TeV in the $h_1$ scenario ($h_2$ scenario under this hierarchy is heavily constrained by the LZ 2024 limits). 
As shown,  the values of $\Delta$ increase monotonically with $\mu_{\rm tot}$, which is consistent with the analysis in Sec \ref{EW fine-tuning}.
If $\Delta \leq 40$ is required as a criterion for naturalness,  we obtain an upper bound $\mu_{\text{tot}} \lesssim 500~\text{GeV}$.
Within the mass hierarchy $m_{\tilde{S}} < m_{\tilde{B}} < m_{\tilde{H}}$ with $M_1 = 150~\text{GeV}$,
numerous viable samples satisfy $\Delta \leq 40$, and the smallest value of $\Delta$ is about 5.6 at $\mu_{\rm tot} \simeq 200~\text{GeV}$.
By contrast, for $m_{\tilde{S}} < m_{\tilde{H}} < m_{\tilde{B}}$ with $M_1$= 2 TeV, LHC searches for electroweakinos constrain $\mu_{\rm tot}\geq$ 900 GeV, leading to $\Delta >$ 140 which corresponds to a tuning level worse than 0.7\%.
These results demonstrate that naturalness provides a  discriminant among competing mass hierarchies. All subsequent analyses will be confined to the  hierarchy $m_{\tilde{S}} < m_{\tilde{B}} < m_{\tilde{H}}$  with $M_1 = 150~\text{GeV}$.

\subsection{Naturalness and Tens-of-GeV Dark Matter}

\begin{figure}
\centering
\includegraphics[width=0.5\linewidth]{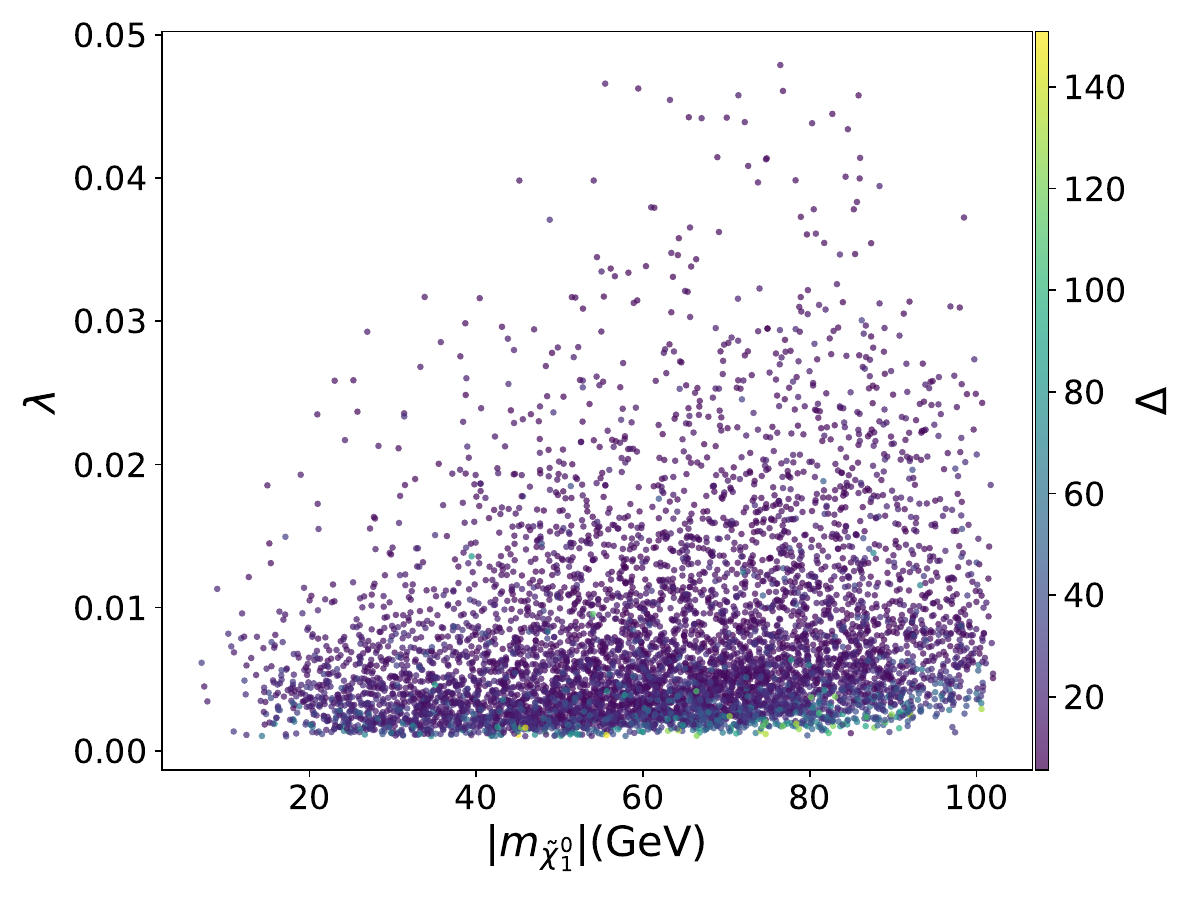}
   \hspace{-0.3cm}
\includegraphics[width=0.5\linewidth]{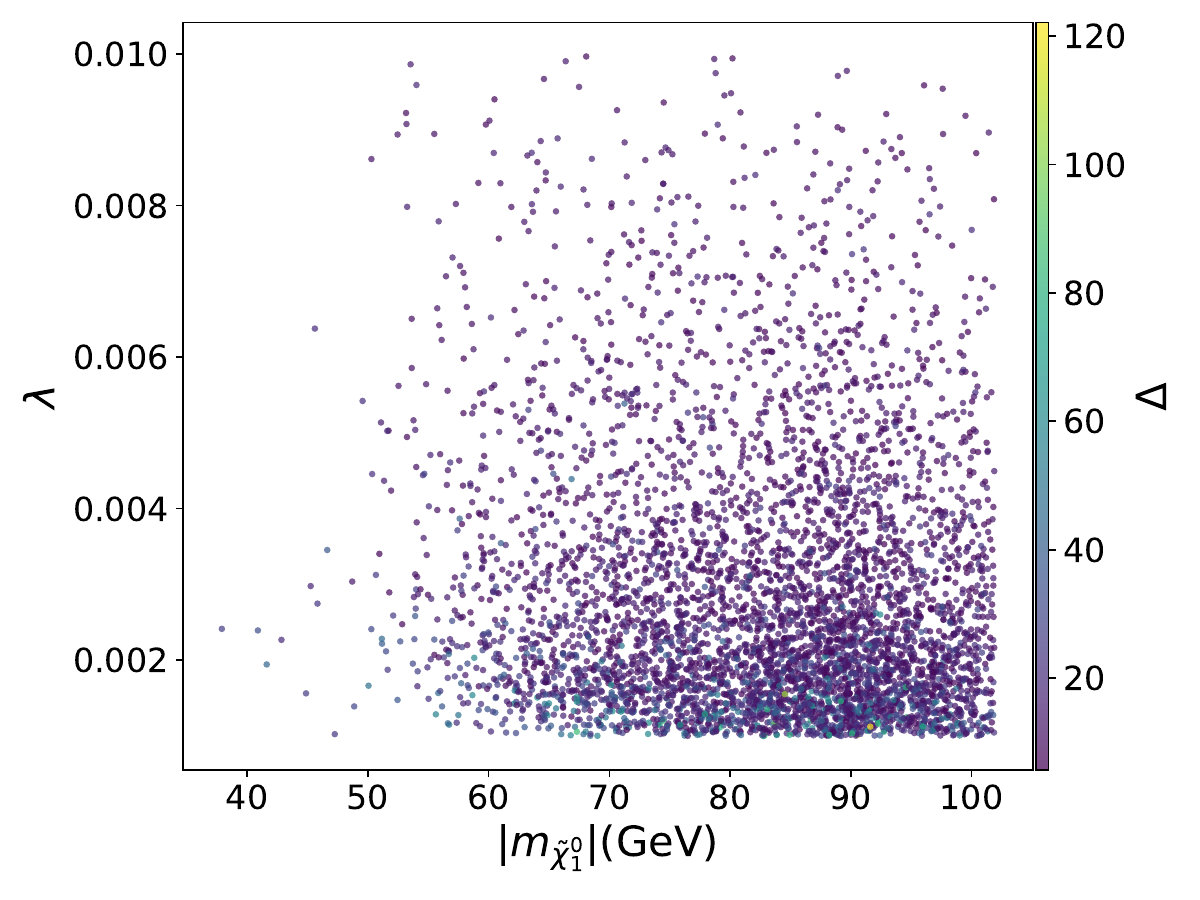}
\caption{$|m_{\tilde{\chi}_1^0}|$ versus $\lambda$ for the $h_1$ scenario (left) and the $h_2$ scenario (right), with $M_1 = 150~\text{GeV}$. The color bar represents the fine-tuning measure $\Delta$.\label{fig2}}
\end{figure}

Within the hierarchy $m_{\tilde{S}} < m_{\tilde{B}} < m_{\tilde{H}}$ with $M_1 = 150~\text{GeV}$, the lightest neutralino $\tilde{\chi}_1^0$ is nearly a pure singlino with $|m_{\tilde{\chi}_1^0}| \simeq |m_{\tilde{S}}|$. 
Figure~\ref{fig2} displays the surviving samples of this hierarchy in the plane of $|m_{\tilde{\chi}_1^0}|$ versus $\lambda$, with the color bar indicating the electroweak fine-tuning measure $\Delta$. 
In the $h_1$ scenario (left panel), the mass of $\tilde{\chi}_1^0$ spans a broad range of 10--100 GeV.
Notably, almost all natural samples with $\Delta \leq 40$ are concentrated within this mass range.
In the $h_2$ scenario (right panel), by contrast, the natural region is shifted toward a heavier DM, $|m_{\tilde{\chi}_1^0}| \gtrsim$ 50 GeV. 
This shift arises from the combined requirements of Higgs physics and the
measured relic density.
Specifically, constraints from exotic Higgs decay searches, together with the kinematic requirement $2|m_{\tilde{\chi}_1^0}| \gtrsim m_{h_s} + m_{a_s}$ for the dominant annihilation channel $\tilde{\chi}_1^0\tilde{\chi}_1^0 \to h_s a_s$, favor a spectrum  where $m_{h_s} > |m_{\tilde{\chi}_1^0}|\gtrsim 60~\text{GeV}$ for most of the parameter space \cite{Li:2025qkg}.

The allowed range of $\lambda$ also differs markedly between the $h_1$ and $h_2$ scenarios: $0 < \lambda \lesssim 0.05$ in the $h_1$ case, shrinking considerably to $0 < \lambda \lesssim 0.01$ in the $h_2$ case.
The tighter bound arises from the Higgs sector.
In the $h_2$ scenario the lightest CP-even Higgs boson is singlet-dominated and lies much closer in mass to the SM-like $h$ than in the $h_1$ scenario.
As follows from $V_h^{\rm S}/V_h^{\rm SM}\simeq\mathcal{M}_{S,23}^2/(m_h^2-\mathcal{M}_{S,33}^2)$ \cite{Li:2025qkg}, the SM-like $h$ then acquires a larger singlet admixture, especially when $h_s$ is nearly degenerate with $h$ in mass, and its couplings to the singlet states are correspondingly enhanced.
Since $\mathcal{M}_{S,23}^2 \propto \lambda$, the singlet admixture and the associated deviations in Higgs couplings from SM predictions increase linearly with $\lambda$.
As a result, global fits to Higgs couplings, combined with LHC searches for exotic Higgs decays, impose significantly tighter constraints on $\lambda$ in the $h_2$ scenario.

Furthermore, in both scenarios, the  highly fine-tuned samples ($\Delta > 80$) cluster at the lower boundary of the allowed $\lambda$ region, where the singlet--doublet mixing becomes negligible and the phenomenology converges to the MSSM with a bino-like neutralino.
Consequently, the accumulation of large-$\Delta$ samples at small $\lambda$ directly reflects the fine-tuning inherent to this MSSM-like limit.

\begin{figure}
\centering
\includegraphics[width=0.5\linewidth]{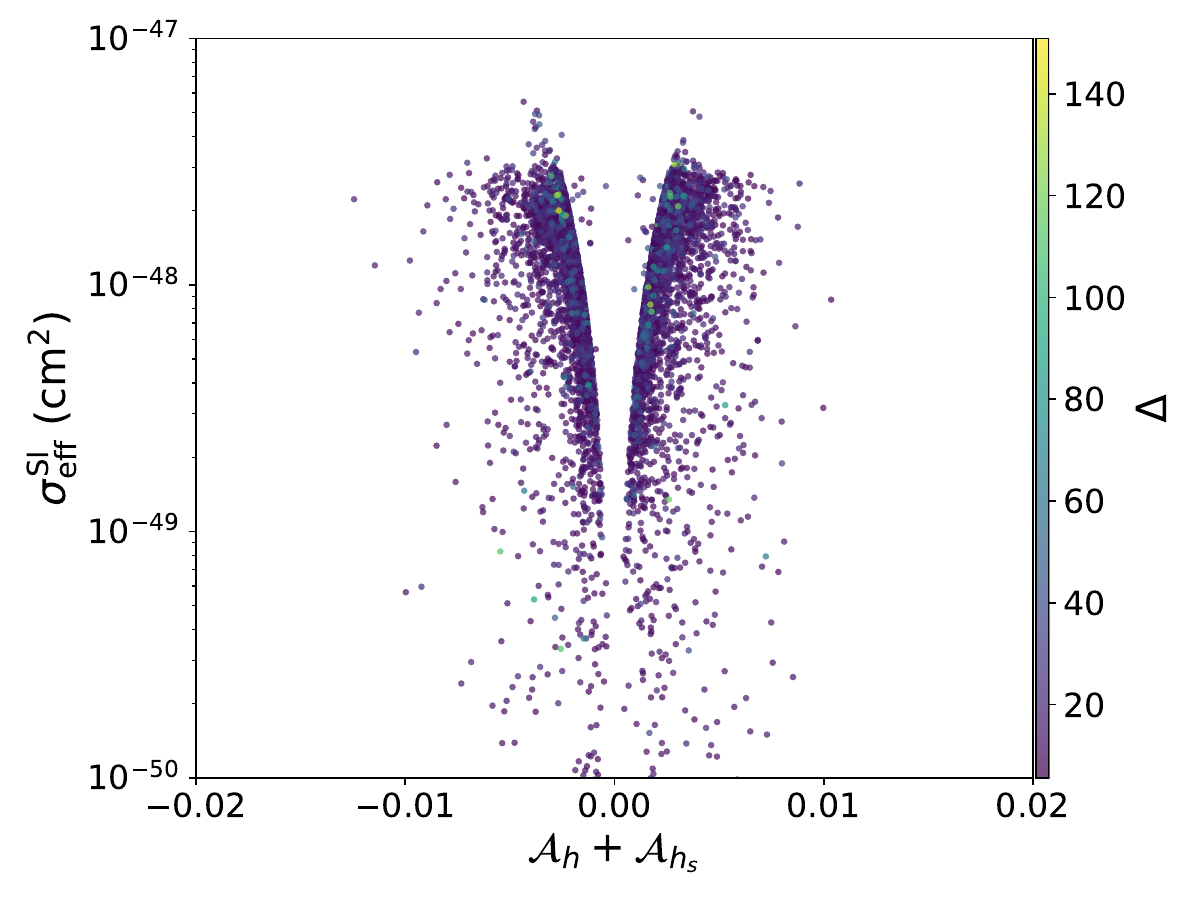}
   \hspace{-0.3cm}
\includegraphics[width=0.5\linewidth]{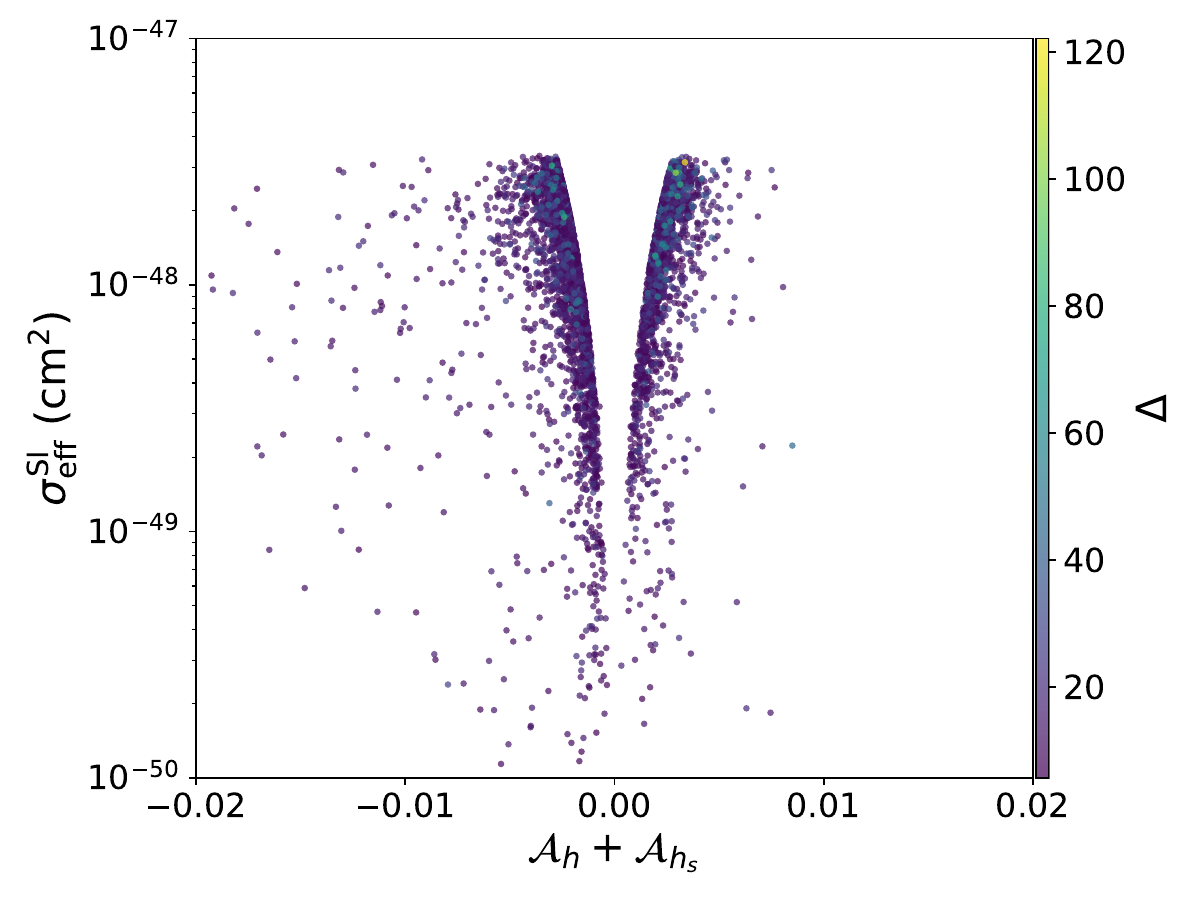}
\caption{Same as Fig.~\ref{fig2}, but showing $\sigma_{\rm eff}^{\rm SI}$ versus the total SI scattering amplitude ($\mathcal{A}_h + \mathcal{A}_{h_s}$) for the $h_1$ scenario (left) and the $h_2$ scenario (right).\label{fig3}}
\end{figure}

\begin{figure}
\centering
\includegraphics[width=0.5\linewidth]{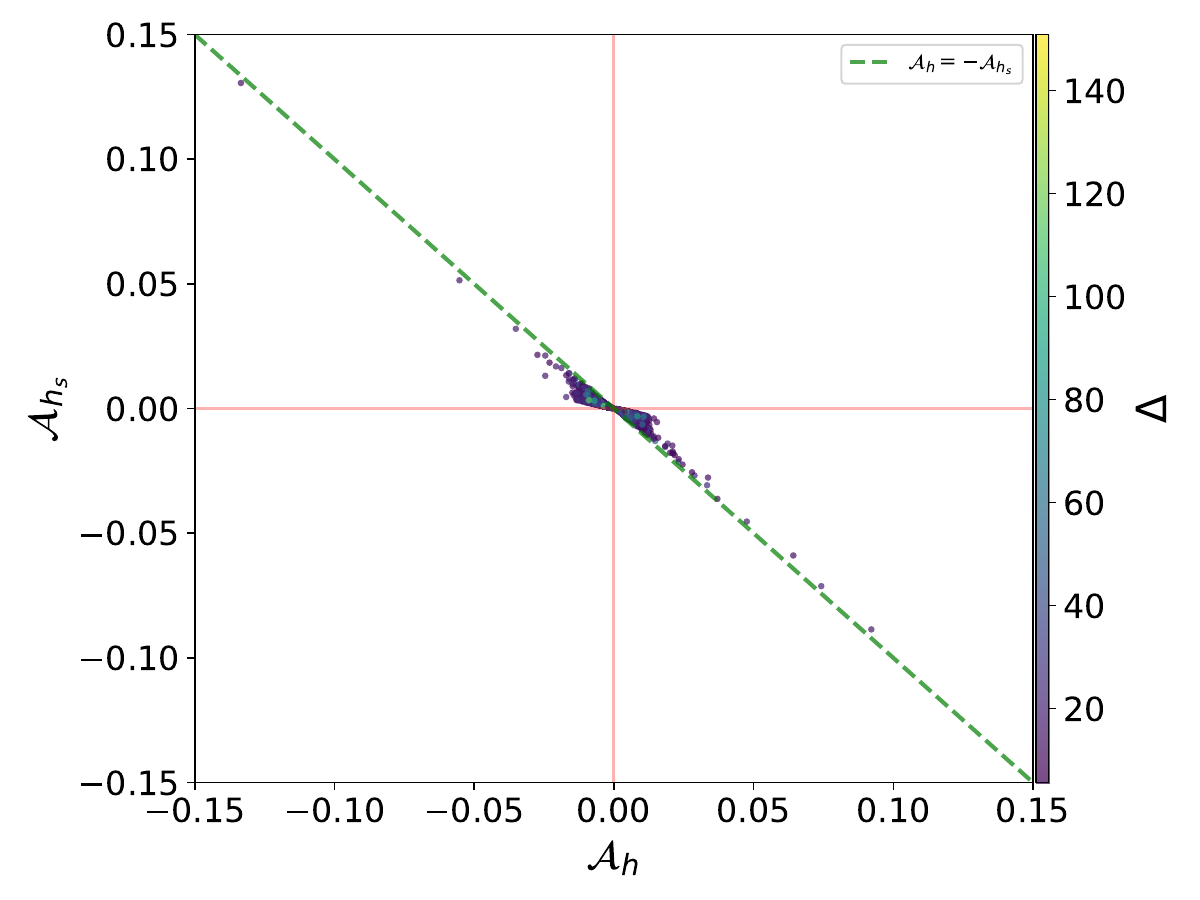}
   \hspace{-0.3cm}
\includegraphics[width=0.5\linewidth]{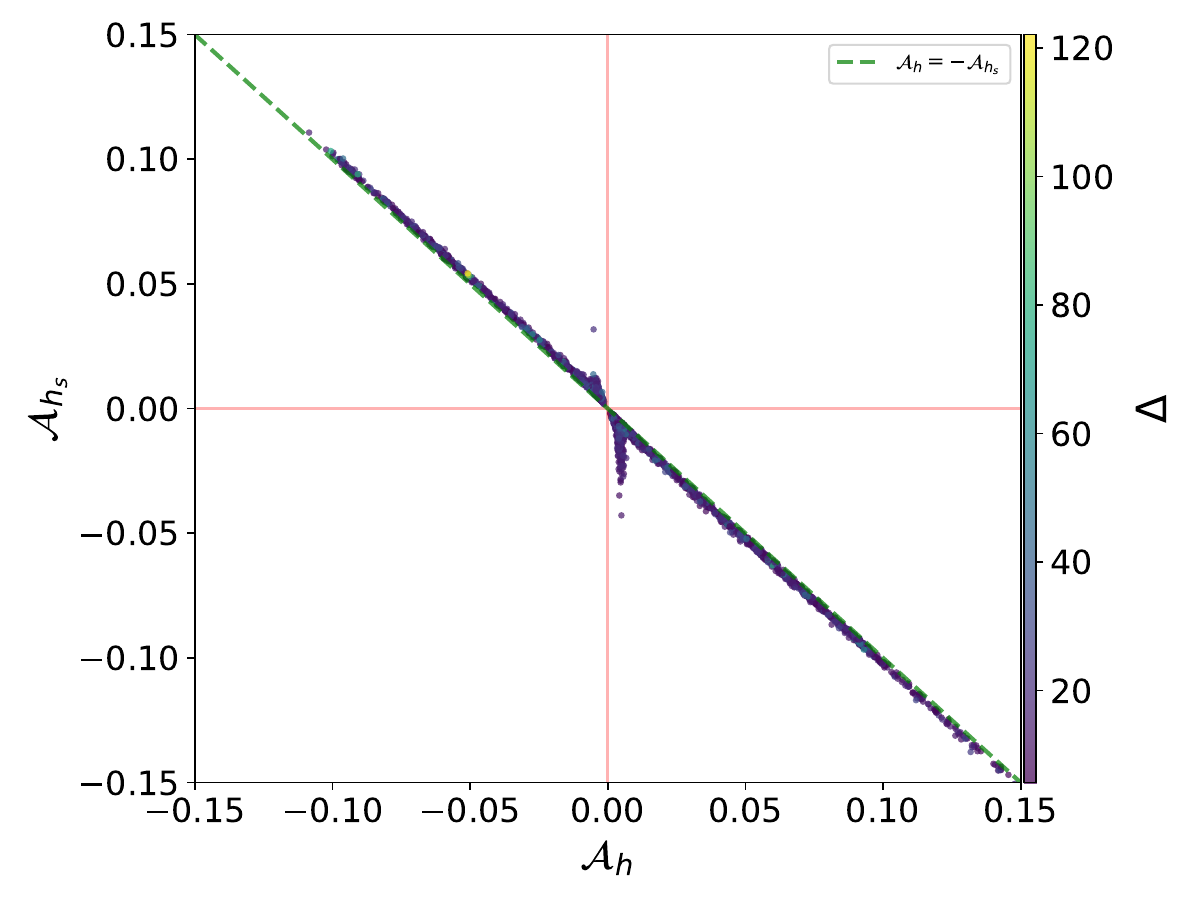}
\caption{Same as Fig.~\ref{fig2}, but showing the individual amplitudes $\mathcal{A}_h$ versus $\mathcal{A}_{h_s}$ for the $h_1$ scenario (left) and the $h_2$ scenario (right).\label{fig4}}
\end{figure}
 
Figure~\ref{fig3} displays the same surviving samples in the plane of $\sigma_{\rm eff}^{\rm SI}$ versus the total SI scattering amplitude ($\mathcal{A}_h + \mathcal{A}_{h_s}$), again color-coded by $\Delta$. 
The characteristic V-shaped pattern reflects that the two amplitudes $\mathcal{A}_h$ and $\mathcal{A}_{h_s}$ interfere destructively.
As $\mathcal{A}_h + \mathcal{A}_{h_s}$ approaches zero, $\sigma_{\rm eff}^{\rm SI}$ reaches $\mathcal{O}(10^{-50})~\text{cm}^2$ in both $h_1$ and $h_2$ scenarios.
Both scenarios exhibit a common phenomenological characteristic: natural parameter regions ($\Delta \leq 40$) correspond to $\sigma_{\rm eff}^{\rm SI}$ between $10^{-50}~\text{cm}^2$ and $10^{-48}~\text{cm}^2$, whereas the highly fine-tuned samples ($\Delta > 80$) cluster near the LZ 2024 exclusion limit.

As shown in Figure \ref{fig4}, the individual amplitudes exhibit different behaviors in the two scenarios.
In the $h_1$ scenario the samples are confined to $|\mathcal{A}_{h,h_s}| \lesssim 0.05$, whereas in the $h_2$ scenario they extend to $|\mathcal{A}_{h,h_s}| \simeq 0.15$ along the anti-diagonal line $\mathcal{A}_h = -\mathcal{A}_{h_s}$.
This enhancement stems from the larger singlet admixture $V_h^{S}$ of the SM-like $h$ and the light-$h_s$ propagator factor $(125/m_{h_s})^2 > 1$, which amplify both amplitudes $\mathcal{A}_h$ and $\mathcal{A}_{h_s}$ in the $h_2$ scenario, as described by Eqs.~(\ref{amplitude}) and (\ref{coupling-chih}).
Since the LZ 2024 limit constrains the total amplitude $|\mathcal{A}_h + \mathcal{A}_{h_s}|$ to a comparable magnitude in both cases, the surviving $h_2$ samples are confined to cluster tightly along the cancellation line $\mathcal{A}_h = -\mathcal{A}_{h_s}$.
Consequently, the suppression of $\sigma_{\rm eff}^{\rm SI}$ in the $h_2$ scenario is highly sensitive to the exact cancellation between $\mathcal{A}_h$ and $\mathcal{A}_{h_s}$.

Since the leading contribution to the SI cross section scales as
$\lambda^2\kappa^2$ in both scenarios, the cancellation mechanism between amplitudes also explains the stringent bound $\lambda \lesssim 0.01$ in the $h_2$ scenario.
The tighter bound $\lambda \lesssim 0.01$ found in the $h_2$ scenario is thus
not only a consequence of the Higgs coupling fits and the exotic Higgs decay
searches, but is also corroborated by the direct-detection constraint.

\subsection{Naturalness and Exotic Higgs Decays}

\begin{figure}
\centering
\includegraphics[width=0.5\linewidth]{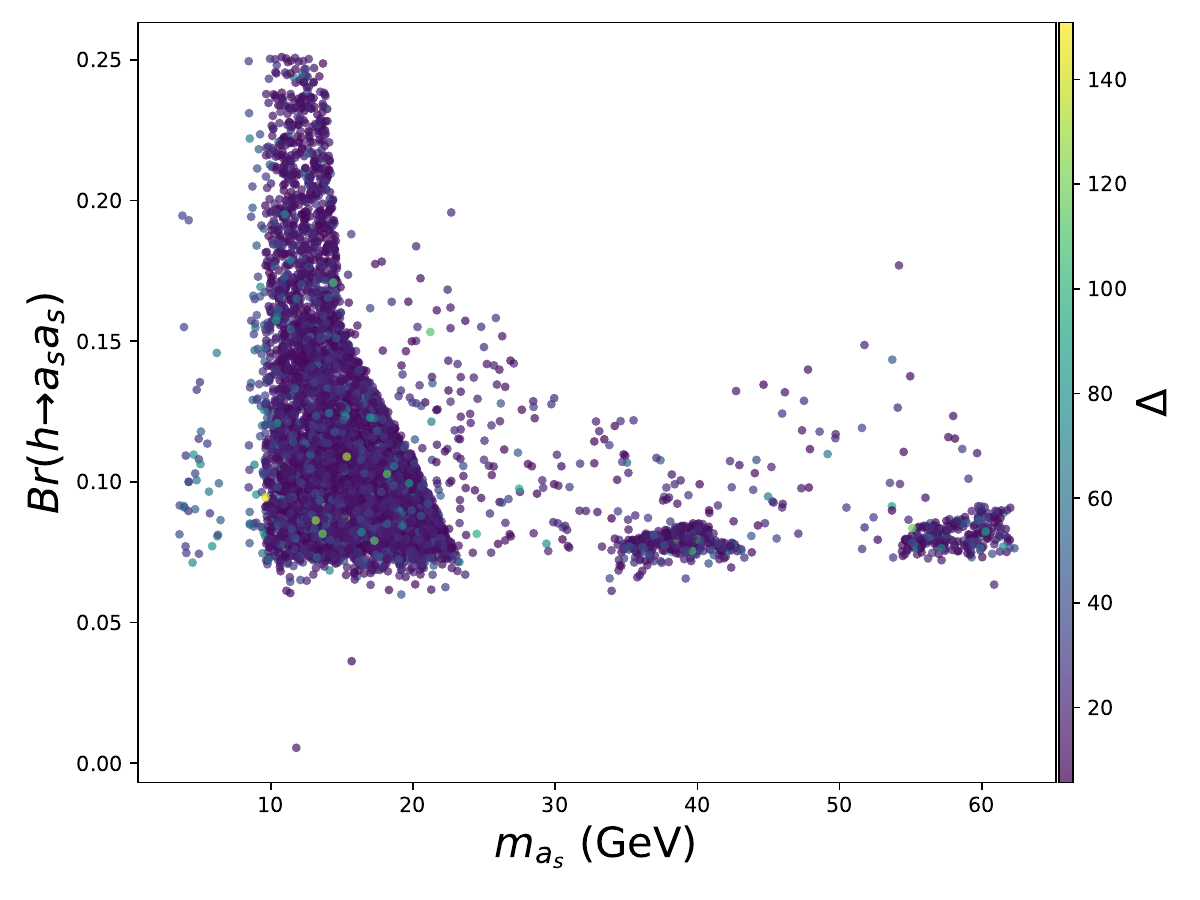}
   \hspace{-0.3cm}
\includegraphics[width=0.5\linewidth]{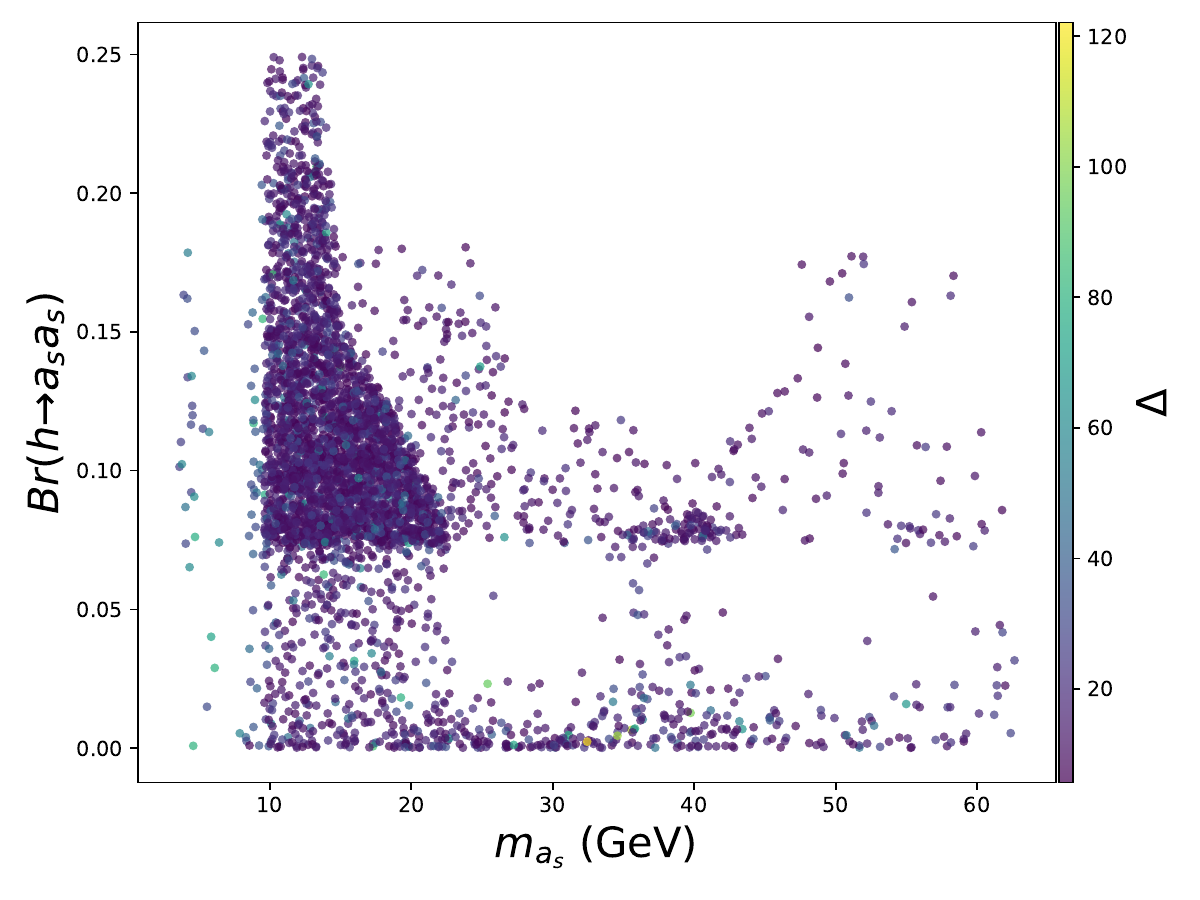}
\caption{Same as Fig.~\ref{fig2}, but showing $\text{Br}(h\to a_s a_s)$ versus $m_{a_s}$ for the samples in the $h_1$ scenario (left panel) and $h_2$ scenario (right panel). \label{fig5}}
\end{figure}

\begin{figure}
\centering
\includegraphics[width=0.5\linewidth]{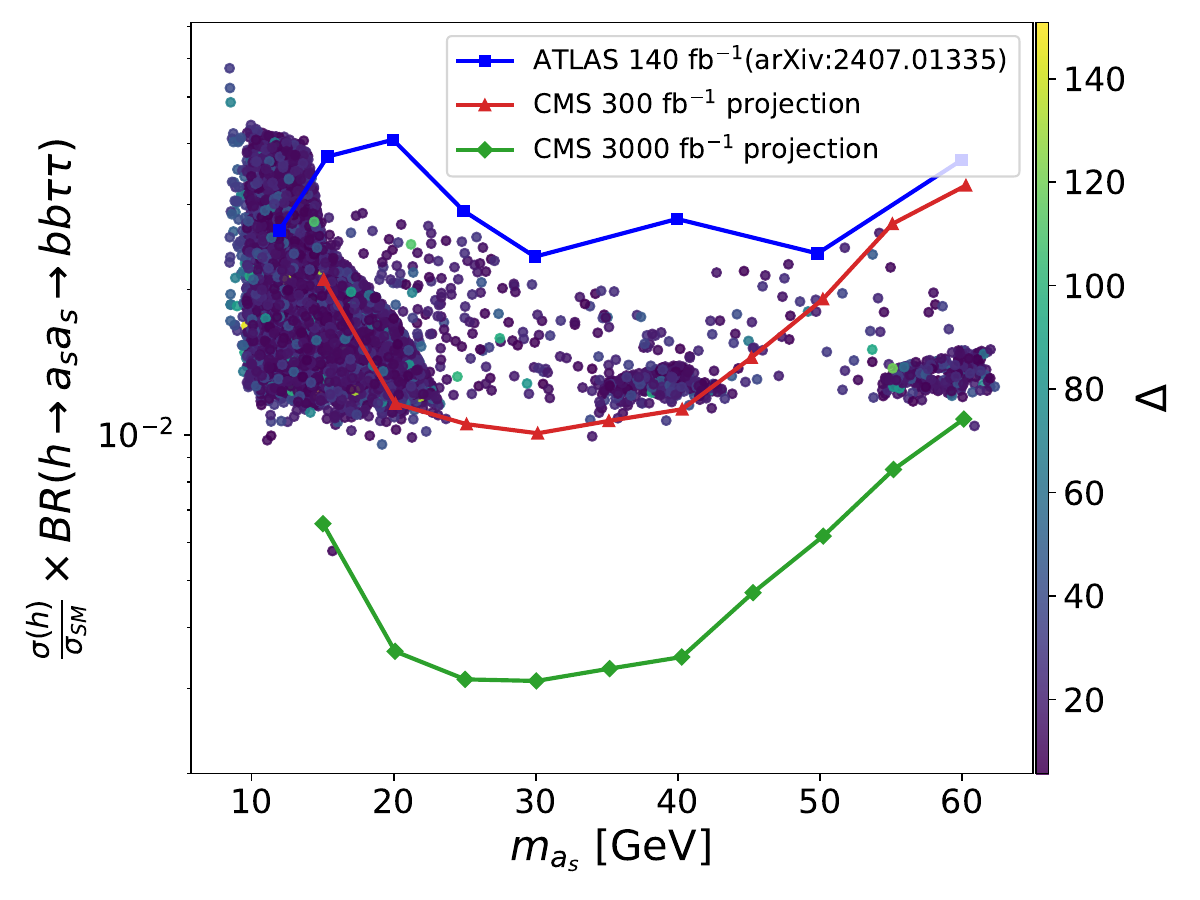}
   \hspace{-0.3cm}
\includegraphics[width=0.5\linewidth]{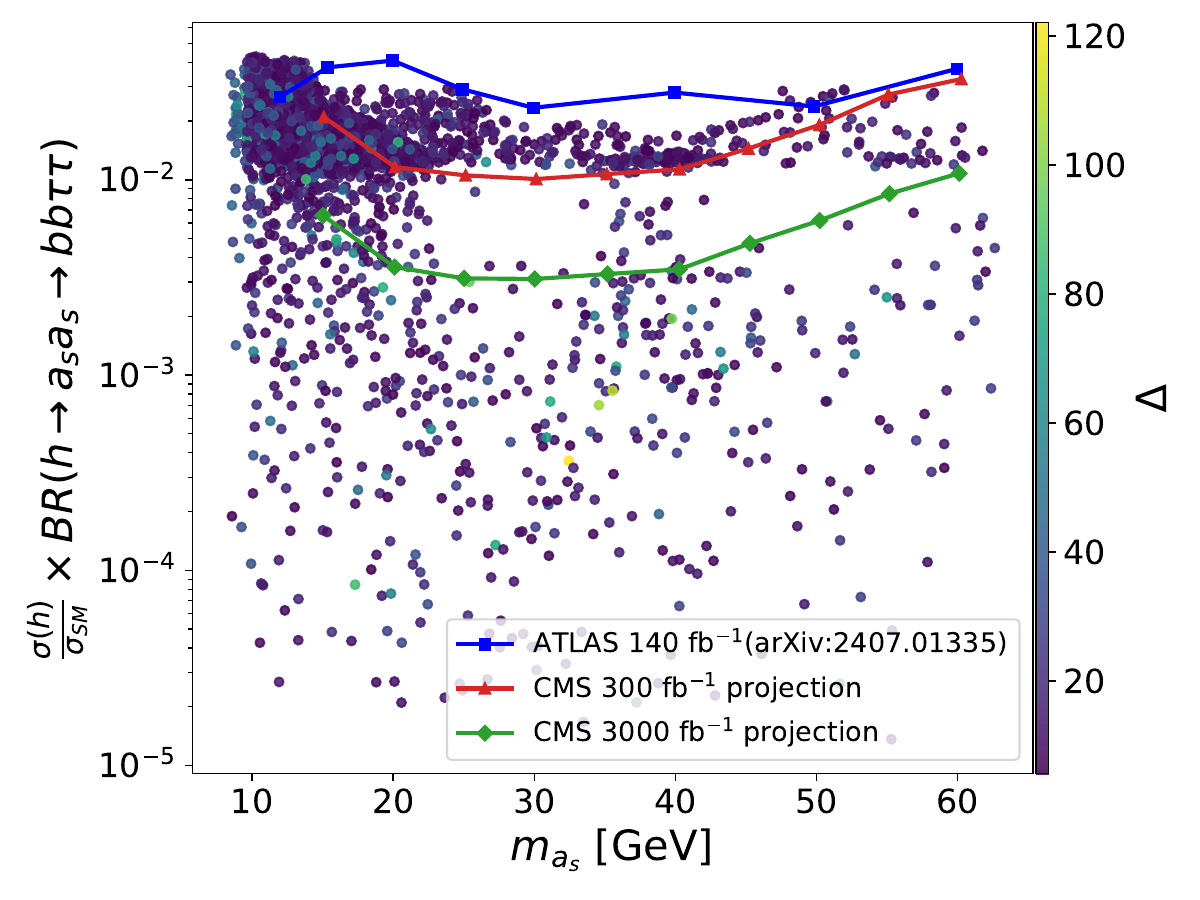}
\caption{Same as Fig.~\ref{fig2}, but showing $\sigma (h)/\sigma_{\rm SM}\times \text{Br}(h\to a_s a_s\to bb\tau\tau)$ versus $m_{a_s}$ for the samples in the $h_1$ scenario (left panel) and the $h_2$ scenario (right panel).\label{fig6}}
\end{figure}

\begin{figure}
\centering
\includegraphics[width=0.5\linewidth]{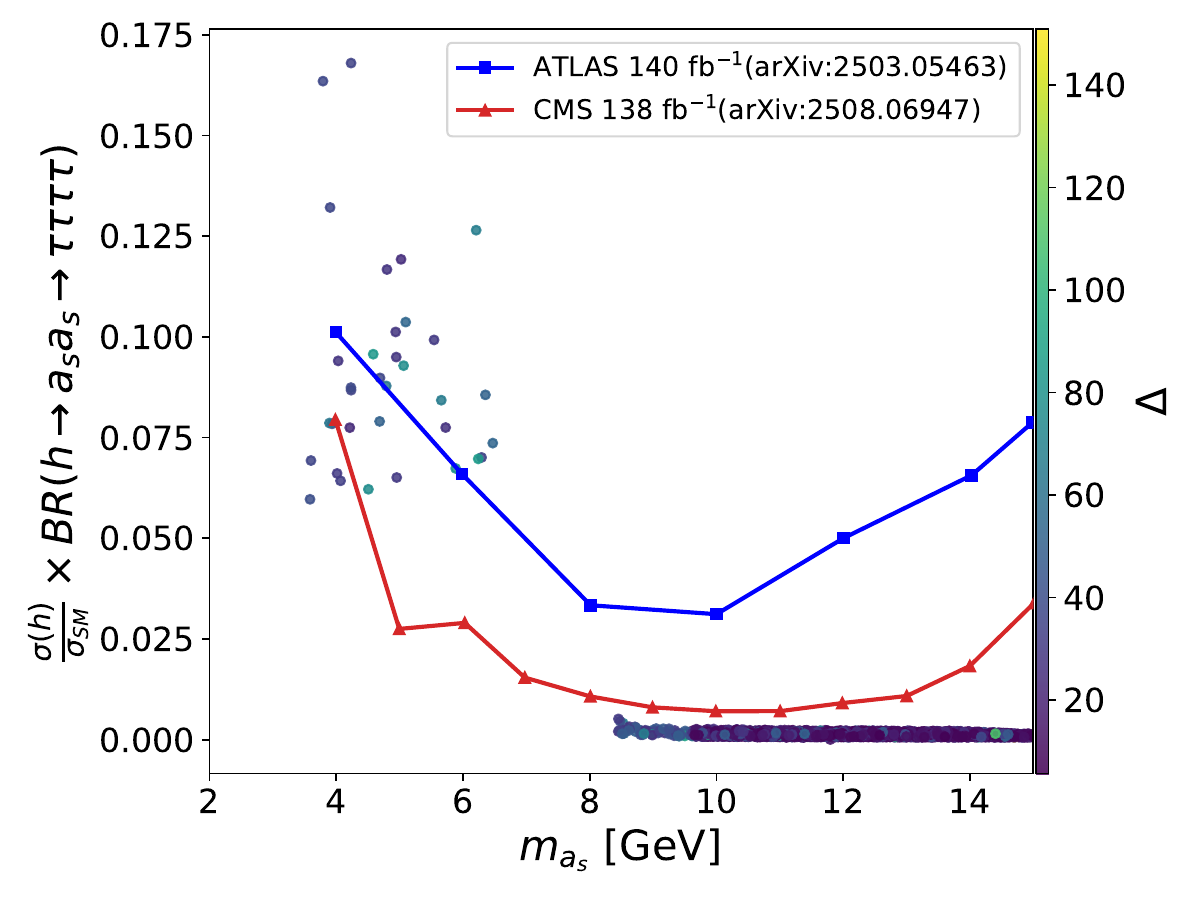}
   \hspace{-0.3cm}
\includegraphics[width=0.5\linewidth]{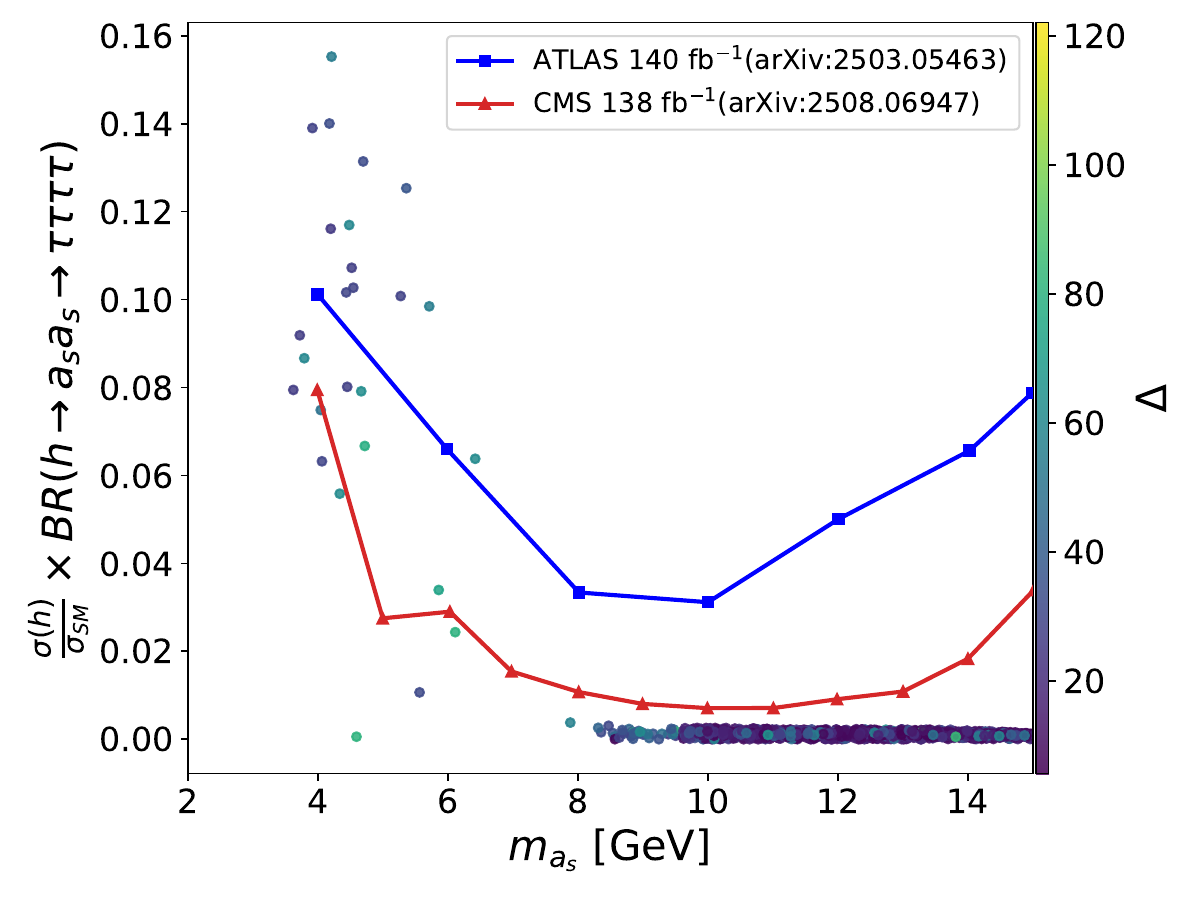}
\caption{Same as Fig.~\ref{fig2}, but showing $\sigma (h)/\sigma_{\rm SM}\times \text{Br}(h\to a_s a_s\to \tau\tau\tau\tau)$ versus $m_{a_s}$ for the samples in the $h_1$ scenario (left panel) and the $h_2$ scenario (right panel).\label{fig7}}
\end{figure}

The natural window with tens-of-GeV DM is also characterized by a distinctive collider signature, that is, the exotic decay of the SM-like Higgs boson into a pair of light pseudoscalar singlets, $h\to a_s a_s$.
Indeed, the vast majority of nature samples ($\Delta \leq 40$) predict a light $a_s$, with $m_{a_s}$ concentrated in the 10--20 GeV window ($78\%$ in the $h_1$
scenario and $70\%$ in the $h_2$ scenario) and about half of them below
15 GeV.
Such a light $a_s$ automatically opens the exotic decay channel $h \to a_s a_s$. 
Figure \ref{fig5} shows the distribution of $\text{Br}(h \to a_s a_s)$ as a function of $m_{a_s}$, with the color bar indicating $\Delta$.
In the natural region $\Delta \leq 40$, the branching ratio reaches $\simeq$ 25\% for $m_{a_s}\lesssim$ 15 GeV, and increases as $m_{a_s}$ decreases. 

Currently, LHC experiments have performed detailed searches for exotic decay $h\to a_s a_s$, focusing on the combine channels such as $4\tau$, $2\mu2\tau$ $4b$, and $bb\tau\tau$, but no significant deviations from the SM expectation have been observed \cite{Aravind:2026jxz, CMS:2024zfv}.
The available data have already established upper limits on $\sigma(h)/\sigma_{\rm SM}(h)\times{\rm BR}(h\to a_sa_s)$, thereby imposing stringent constraints on the GNMSSM parameter space for ${\rm BR}(h\to a_sa_s)\sim 7\%$--$25\%$.
Figure \ref{fig6} shows the 95\% CL upper limit on $\sigma (h)/\sigma_{\rm SM}(h)\times \text{Br}(h\to a_s a_s\to bb\tau\tau)$ obtained by ATLAS with an integrated luminosity of 140 fb$^{-1}$ \cite{ATLAS2024}, together with the CMS projected sensitivities for 300 fb$^{-1}$ and 3000 fb$^{-1}$ \cite{Cepeda:2019klc}.
The ATLAS 140 fb$^{-1}$ limits, which apply to 12 GeV $\leq m_{a_s} \leq$ 60 GeV, exclude the upper envelope of the predicted distributions in both $h_1$ and $h_2$ scenarios, primarily in 12 GeV $\lesssim m_{a_s}\lesssim$ 15 GeV, where $\text{BR}(h\to a_sa_s)$ is largest, as shown in Figure \ref{fig5}.
Approximately $12\%$ ($10\%$) of the samples with 12 GeV $\leq m_{a_s} \leq 15.5$~GeV are excluded in the $h_1$ ($h_2$) scenario.
The 300 fb$^{-1}$ projection would exclude about half of the samples in the visible band, and the 3000 fb$^{-1}$ reach lies below essentially all scanned points in the $h_1$ scenario.
However, in the $h_2$ scenario, there are still some viable samples remaining despite the HL-LHC $bb\tau\tau$ searches, due to the reduced $\sigma(h)/\sigma_{\rm SM}(h)$ and/or the diluted $\text{BR}(a_s\to b\bar b)$.
And it may be probed by complementary channels such as $4b$, $4\tau/2\mu2\tau$ or $bb\mu\mu$. 
Figure~\ref{fig6} also shows that the largest and most testable signals come mainly from the most natural samples ($\Delta\lesssim 20$), therefore, a future null result in the $bb\tau\tau$ channel would push the GNMSSM with tens-of-GeV DM toward larger fine-tuning.

To probe $m_{a_s}$ below the reach of the $bb\tau\tau$ searches, Figure~\ref{fig7} displays the scanned samples with $m_{a_s}\leq 15$ GeV in the plane of $\sigma(h)/\sigma_{\rm SM}(h)\times{\rm BR}(h\to a_sa_s\to \tau^+\tau^-\tau^+\tau^-)$ versus $m_{a_s}$, together with the observed 95\% CL upper limits from ATLAS ($140~{\rm fb}^{-1}$) \cite{ATLAS:2025qyn} and CMS ($138~{\rm fb}^{-1}$) \cite{CMS:2025hjt}, both covering $m_{a_s}$ in the range of 4--15 GeV.
Below the $2m_b$ threshold ($m_{a_s}\lesssim 8.4$~GeV), the decay $a_s\to b\bar b$ is kinematically closed, ${\rm BR}(a_s\to\tau\tau)\simeq 0.96$, and the $4\tau$ final state is the only sensitive probe.
The current limits, $(\sigma(h)/\sigma_{\rm SM}(h)) \times{\rm BR}(h\to a_sa_s\to 4\tau)=0.03$--$0.10$ (ATLAS) and $0.007$--$0.079$ (CMS), exclude most of the samples with $m_{a_s}\lesssim 8.4$ GeV in the $h_1$ ($h_2$) scenario.
Above $2m_b$, ${\rm BR}(a_s\to\tau\tau)$ drops to $\simeq 0.1$ and the $4\tau$ rate falls by more than an order of magnitude, while the $bb\tau\tau$ searches lose sensitivity below $12$ GeV.
The mass window with $8.4\lesssim m_{a_s}\lesssim 12$ GeV, which contains a significant fraction of the natural samples, thus remains uncovered by current direct searches and calls for dedicated low-mass strategies at the HL-LHC, such as soft-$b$ tagging or boosted $\tau$-pair reconstruction.

Therefore, if future HL-LHC analyses or $e^+e^-$ colliders (CEPC, FCC-ee, ILC) observe the signal of exotic decay $h\to a_sa_s$, this would not only point to the GNMSSM but also specifically favor the natural mass hierarchy $m_{\tilde{S}}<m_{\tilde{B}}<m_{\tilde{H}}$ with $\mu_{\rm tot}\sim$ 200--400 GeV. 
In contrast, the absence of a signal would disfavor the natural window and push the parameter space toward the highly fine-tuned hierarchy
$m_{\tilde{S}}<m_{\tilde{H}}<m_{\tilde{B}}$, or rule out the GNMSSM as a natural
light-DM framework altogether.

\section{Conclusion}
In this work, we  investigated the interplay between electroweak-scale fine-tuning, tens-of-GeV singlino-dominated DM, and exotic decays of the SM-like Higgs boson in GNMSSM.
Extending the previous DM-oriented analysis of \cite{Li:2025qkg} and the Higgs-only study of \cite{Heng:2025wng}, we have established a direct connection between these three aspects and identified a natural region of the parameter space that simultaneously satisfies the LZ 2024 direct-detection limits and predicts a distinctive collider signature.
The GNMSSM admits a natural window characterized by the mass hierarchy $m_{\tilde{S}}<m_{\tilde{B}}<m_{\tilde{H}}$ with the higgsino mass parameter $\mu_{\rm tot}\in [200, 500]$ GeV.
For natural samples with $\Delta \leq 40$, the branching ratio of the SM-like Higgs boson into a pair of singlet-dominated pseudoscalars ($\text{Br}(h\to a_s a_s)$) reaches approximately 25\% for $m_{a_s}\lesssim$ 15 GeV, and $m_{a_s}$ concentrates mostly between 10 and 20 GeV. 
This $\mathcal{O}(10\%)$ branching ratio is a robust prediction of the natural region, rather than a fine-tuned coincidence.
The current upper limits on $\sigma(h)/\sigma_{\rm SM}(h)\times{\rm BR}(h\to a_sa_s\to bb\tau\tau)$ severely constrain the GNMSSM parameter space, particularly in the region where ${\rm BR}(h\to a_sa_s)\sim 7\%$--$25\%$.
Specifically, about $12\%$ ($10\%$) of the samples with 12 GeV $\leq m_{a_s} \leq 15.5$~GeV are excluded by the ATLAS 140 fb$^{-1}$ limits in the $h_1$ ($h_2$) scenario, while some viable samples still remain despite HL-LHC $bb\tau\tau$ searches in the $h_2$ scenario.

In summary, our work connects DM physics, naturalness, and collider phenomenology wihtin a unified framework. It demonstrates that the GNMSSM not only naturally accommodates a light singlino-dominated DM candidate consistent with current direct-detection limits, but also predicts a distinctive and testable exotic Higgs signature. This makes the GNMSSM a compelling target for future experimental searches and provides a concrete handle for probing the connection between naturalness and dark matter at the electroweak scale.

\section*{Acknowledgement}
This work was supported by the National Natural Science Foundation of China (Grant No. 12505122) and the Natural Science Foundation of Henan Province (Grant No. 252300423511).

\bibliographystyle{CitationStyle}
\bibliography{sample}

@article{Baer:2012cf,
    author = "Baer, Howard and Barger, Vernon and Huang, Peisi and Mickelson, Dan and Mustafayev, Azar and Tata, Xerxes",
    title = "{Radiative natural supersymmetry: Reconciling electroweak fine-tuning and the Higgs boson mass}",
    eprint = "1212.2655",
    archivePrefix = "arXiv",
    primaryClass = "hep-ph",
    doi = "10.1103/PhysRevD.87.115028",
    journal = "Phys. Rev. D",
    volume = "87",
    number = "11",
    pages = "115028",
    year = "2013"
}

@article{Baer:2012uy,
    author = "Baer, Howard and Barger, Vernon and Huang, Peisi and Tata, Xerxes",
    title = "{Natural Supersymmetry: LHC, dark matter and ILC searches}",
    eprint = "1203.5539",
    archivePrefix = "arXiv",
    primaryClass = "hep-ph",
    reportNumber = "UH-511-1190-12",
    doi = "10.1007/JHEP05(2012)109",
    journal = "JHEP",
    volume = "05",
    pages = "109",
    year = "2012"
}

@article{Barbieri:1987fn,
    author = "Barbieri, Riccardo and Giudice, G. F.",
    title = "{Upper Bounds on Supersymmetric Particle Masses}",
    reportNumber = "CERN-TH-4825/87",
    doi = "10.1016/0550-3213(88)90171-X",
    journal = "Nucl. Phys. B",
    volume = "306",
    pages = "63--76",
    year = "1988"
}

@article{Ellis:1986yg,
    author = "Ellis, John R. and Enqvist, K. and Nanopoulos, Dimitri V. and Zwirner, F.",
    title = "{Observables in Low-Energy Superstring Models}",
    reportNumber = "CERN-TH-4350-86",
    doi = "10.1142/S0217732386000105",
    journal = "Mod. Phys. Lett. A",
    volume = "1",
    pages = "57",
    year = "1986"
}

@article{Pospelov:2007mp,
    author = "Pospelov, Maxim and Ritz, Adam and Voloshin, Mikhail B.",
    title = "{Secluded WIMP Dark Matter}",
    eprint = "0711.4866",
    archivePrefix = "arXiv",
    primaryClass = "hep-ph",
    doi = "10.1016/j.physletb.2008.02.052",
    journal = "Phys. Lett. B",
    volume = "662",
    pages = "53--61",
    year = "2008"
}

@article{LZ:2024zvo,
    author = "Aalbers, J. and others",
    collaboration = "LZ",
    title = "{Dark Matter Search Results from 4.2 Tonne-Years of Exposure of the LUX-ZEPLIN (LZ) Experiment}",
    eprint = "2410.17036",
    archivePrefix = "arXiv",
    primaryClass = "hep-ex",
    reportNumber = "FERMILAB-PUB-24-0796-V",
    doi = "10.1103/4dyc-z8zf",
    journal = "Phys. Rev. Lett.",
    volume = "135",
    number = "1",
    pages = "011802",
    year = "2025"
}

@article{Arcadi:2017kky,
    author = "Arcadi, Giorgio and Dutra, Ma{\'\i}ra and Ghosh, Pradipta and Lindner, Manfred and Mambrini, Yann and Pierre, Mathias and Profumo, Stefano and Queiroz, Farinaldo S.",
    title = "{The waning of the WIMP? A review of models, searches, and constraints}",
    eprint = "1703.07364",
    archivePrefix = "arXiv",
    primaryClass = "hep-ph",
    doi = "10.1140/epjc/s10052-018-5662-y",
    journal = "Eur. Phys. J. C",
    volume = "78",
    number = "3",
    pages = "203",
    year = "2018"
}

@article{Arcadi:2024ukq,
    author = "Arcadi, Giorgio and Cabo-Almeida, David and Dutra, Ma{\'\i}ra and Ghosh, Pradipta and Lindner, Manfred and Mambrini, Yann and Neto, Jacinto P. and Pierre, Mathias and Profumo, Stefano and Queiroz, Farinaldo S.",
    title = "{The Waning of the WIMP: Endgame?}",
    eprint = "2403.15860",
    archivePrefix = "arXiv",
    primaryClass = "hep-ph",
    doi = "10.1140/epjc/s10052-024-13672-y",
    journal = "Eur. Phys. J. C",
    volume = "85",
    number = "2",
    pages = "152",
    year = "2025"
}

@article{Li:2025qkg,
    author = "Li, Fei and Cao, Junjie",
    title = "{Natural realization of tens-of-GeV dark matter in the general NMSSM}",
    eprint = "2511.06329",
    archivePrefix = "arXiv",
    primaryClass = "hep-ph",
    doi = "10.1103/x7j7-5mtq",
    journal = "Phys. Rev. D",
    volume = "113",
    number = "11",
    pages = "115002",
    year = "2026"
}

@article{Baer:2013gva,
    author = "Baer, Howard and Barger, Vernon and Mickelson, Dan",
    title = "{How conventional measures overestimate electroweak fine-tuning in supersymmetric theory}",
    eprint = "1309.2984",
    archivePrefix = "arXiv",
    primaryClass = "hep-ph",
    reportNumber = "OU-HEP-130931",
    doi = "10.1103/PhysRevD.88.095013",
    journal = "Phys. Rev. D",
    volume = "88",
    number = "9",
    pages = "095013",
    year = "2013"
}

@article{Cao:2018rix,
    author = "Cao, Junjie and He, Yangle and Shang, Liangliang and Zhang, Yang and Zhu, Pengxuan",
    title = "{Current status of a natural NMSSM in light of LHC 13 TeV data and XENON-1T results}",
    eprint = "1810.09143",
    archivePrefix = "arXiv",
    primaryClass = "hep-ph",
    reportNumber = "CoEPP-MN-18-8",
    doi = "10.1103/PhysRevD.99.075020",
    journal = "Phys. Rev. D",
    volume = "99",
    number = "7",
    pages = "075020",
    year = "2019"
}

@article{Heng:2025wng,
    author = "Heng, Zhaoxia and Li, Zehan and Zhou, Haijing",
    title = "{Constraining the light Higgs bosons in the GNMSSM with recent Higgs data}",
    eprint = "2503.04023",
    archivePrefix = "arXiv",
    primaryClass = "hep-ph",
    doi = "10.1016/j.cjph.2026.02.009",
    journal = "Chin. J. Phys.",
    volume = "101",
    pages = "267--281",
    year = "2026"
}

@article{Ellwanger:2009dp,
    author = "Ellwanger, Ulrich and Hugonie, Cyril and Teixeira, Ana M.",
    title = "{The Next-to-Minimal Supersymmetric Standard Model}",
    eprint = "0910.1785",
    archivePrefix = "arXiv",
    primaryClass = "hep-ph",
    reportNumber = "LPT-ORSAY-09-76, CFTP-09-032, LPTA-09-066",
    doi = "10.1016/j.physrep.2010.07.001",
    journal = "Phys. Rept.",
    volume = "496",
    pages = "1--77",
    year = "2010"
}

@article{Ellwanger:1983mg,
    author = "Ellwanger, U.",
    title = "{NONRENORMALIZABLE INTERACTIONS FROM SUPERGRAVITY, QUANTUM CORRECTIONS AND EFFECTIVE LOW-ENERGY THEORIES}",
    doi = "10.1016/0370-2693(83)90557-9",
    journal = "Phys. Lett. B",
    volume = "133",
    pages = "187--191",
    year = "1983"
}

@article{Abel:1996cr,
    author = "Abel, S. A.",
    title = "{Destabilizing divergences in the NMSSM}",
    eprint = "hep-ph/9609323",
    archivePrefix = "arXiv",
    reportNumber = "ULB-TH-96-16",
    doi = "10.1016/S0550-3213(96)00470-1",
    journal = "Nucl. Phys. B",
    volume = "480",
    pages = "55--72",
    year = "1996"
}

@article{Kolda:1998rm,
    author = "Kolda, Christopher F. and Pokorski, Stefan and Polonsky, Nir",
    title = "{Stabilized singlets in supergravity as a source of the mu - parameter}",
    eprint = "hep-ph/9803310",
    archivePrefix = "arXiv",
    reportNumber = "IASSNS-HEP-97-137, CERN-TH-98-75, RU-97-97",
    doi = "10.1103/PhysRevLett.80.5263",
    journal = "Phys. Rev. Lett.",
    volume = "80",
    pages = "5263--5266",
    year = "1998"
}

@article{Panagiotakopoulos:1998yw,
    author = "Panagiotakopoulos, C. and Tamvakis, K.",
    title = "{Stabilized NMSSM without domain walls}",
    eprint = "hep-ph/9809475",
    archivePrefix = "arXiv",
    doi = "10.1016/S0370-2693(98)01493-2",
    journal = "Phys. Lett. B",
    volume = "446",
    pages = "224--227",
    year = "1999"
}

@article{Ross:2011xv,
    author = "Ross, Graham G. and Schmidt-Hoberg, Kai",
    title = "{The Fine-Tuning of the Generalised NMSSM}",
    eprint = "1108.1284",
    archivePrefix = "arXiv",
    primaryClass = "hep-ph",
    reportNumber = "OUTP-11-48P",
    doi = "10.1016/j.nuclphysb.2012.05.007",
    journal = "Nucl. Phys. B",
    volume = "862",
    pages = "710--719",
    year = "2012"
}

@article{Lee:2010gv,
    author = "Lee, Hyun Min and Raby, Stuart and Ratz, Michael and Ross, Graham G. and Schieren, Roland and Schmidt-Hoberg, Kai and Vaudrevange, Patrick K. S.",
    title = "{A unique $\mathbb{Z}_4^R$ symmetry for the MSSM}",
    eprint = "1009.0905",
    archivePrefix = "arXiv",
    primaryClass = "hep-ph",
    reportNumber = "TUM-HEP-770-10, LMU-ASC-64-10, OHSTPY-HEP-T-10-003, CERN-PH-TH-2010-193, OUTP-10-24P",
    doi = "10.1016/j.physletb.2010.10.038",
    journal = "Phys. Lett. B",
    volume = "694",
    pages = "491--495",
    year = "2011"
}

@article{Lee:2011dya,
    author = "Lee, Hyun Min and Raby, Stuart and Ratz, Michael and Ross, Graham G. and Schieren, Roland and Schmidt-Hoberg, Kai and Vaudrevange, Patrick K. S.",
    title = "{Discrete R symmetries for the MSSM and its singlet extensions}",
    eprint = "1102.3595",
    archivePrefix = "arXiv",
    primaryClass = "hep-ph",
    reportNumber = "TUM-HEP-793-11, LMU-ASC-06-11, OHSTPY-HEP-T-11-001, CERN-PH-TH-2011-022, OUTP-11-33P",
    doi = "10.1016/j.nuclphysb.2011.04.009",
    journal = "Nucl. Phys. B",
    volume = "850",
    pages = "1--30",
    year = "2011"
}

@article{Ross:2012nr,
    author = "Ross, Graham G. and Schmidt-Hoberg, Kai and Staub, Florian",
    title = "{The Generalised NMSSM at One Loop: Fine Tuning and Phenomenology}",
    eprint = "1205.1509",
    archivePrefix = "arXiv",
    primaryClass = "hep-ph",
    reportNumber = "OUTP-12-06P, BONN-TH-2012-04",
    doi = "10.1007/JHEP08(2012)074",
    journal = "JHEP",
    volume = "08",
    pages = "074",
    year = "2012"
}

@article{Meng:2024lmi,
    author = "Meng, Lei and Cao, Junjie and Li, Fei and Yang, Shenshen",
    title = "{Dark Matter physics in general NMSSM}",
    eprint = "2405.07036",
    archivePrefix = "arXiv",
    primaryClass = "hep-ph",
    doi = "10.1007/JHEP08(2024)212",
    journal = "JHEP",
    volume = "08",
    pages = "212",
    year = "2024"
}

@article{Ellwanger:2004xm,
    author = "Ellwanger, Ulrich and Gunion, John F. and Hugonie, Cyril",
    title = "{NMHDECAY: A Fortran code for the Higgs masses, couplings and decay widths in the NMSSM}",
    eprint = "hep-ph/0406215",
    archivePrefix = "arXiv",
    reportNumber = "LPT-ORSAY-04-32, UCD-04-22, IFIC-04-33",
    doi = "10.1088/1126-6708/2005/02/066",
    journal = "JHEP",
    volume = "02",
    pages = "066",
    year = "2005"
}

@article{Ellwanger:2005dv,
    author = "Ellwanger, Ulrich and Hugonie, Cyril",
    title = "{NMHDECAY 2.0: An Updated program for sparticle masses, Higgs masses, couplings and decay widths in the NMSSM}",
    eprint = "hep-ph/0508022",
    archivePrefix = "arXiv",
    reportNumber = "LPT-ORSAY-05-52",
    doi = "10.1016/j.cpc.2006.04.004",
    journal = "Comput. Phys. Commun.",
    volume = "175",
    pages = "290--303",
    year = "2006"
}

@article{Planck:2018vyg,
    author = "Aghanim, N. and others",
    collaboration = "Planck",
    title = "{Planck 2018 results. VI. Cosmological parameters}",
    eprint = "1807.06209",
    archivePrefix = "arXiv",
    primaryClass = "astro-ph.CO",
    doi = "10.1051/0004-6361/201833910",
    journal = "Astron. Astrophys.",
    volume = "641",
    pages = "A6",
    year = "2020",
    note = "[Erratum: Astron.Astrophys. 652, C4 (2021)]"
}

@article{Bahl:2022igd,
    author = {Bahl, Henning and Biek{\"o}tter, Thomas and Heinemeyer, Sven and Li, Cheng and Paasch, Steven and Weiglein, Georg and Wittbrodt, Jonas},
    title = "{HiggsTools: BSM scalar phenomenology with new versions of HiggsBounds and HiggsSignals}",
    eprint = "2210.09332",
    archivePrefix = "arXiv",
    primaryClass = "hep-ph",
    doi = "10.1016/j.cpc.2023.108803",
    journal = "Comput. Phys. Commun.",
    volume = "291",
    pages = "108803",
    year = "2023"
}

@article{Cao:2024axg,
    author = "Cao, Junjie and Jia, Xinglong and Lian, Jingwei",
    title = "{Unified interpretation of the muon g-2 anomaly, the 95~GeV diphoton, and bb{\textasciimacron} excesses in the general next-to-minimal supersymmetric standard model}",
    eprint = "2402.15847",
    archivePrefix = "arXiv",
    primaryClass = "hep-ph",
    doi = "10.1103/PhysRevD.110.115039",
    journal = "Phys. Rev. D",
    volume = "110",
    number = "11",
    pages = "115039",
    year = "2024"
}

@inproceedings{Aravind:2026jxz,
    author = "Aravind, Anagha",
    title = "{Rare decays of the SM Higgs to light pseudoscalars in the CMS Experiment}",
    booktitle = "{32nd International Symposium on Lepton Photon Interactions at High Energies}: {Lepton-Photon 2025}",
    eprint = "2605.01714",
    archivePrefix = "arXiv",
    primaryClass = "hep-ex",
    reportNumber = "CMS CR-2025/262",
    month = "5",
    year = "2026"
}

@article{CMS:2024zfv,
    author = "Hayrapetyan, Aram and others",
    collaboration = "CMS",
    title = "{Search for the decay of the Higgs boson to a pair of light pseudoscalar bosons in the final state with four bottom quarks in proton-proton collisions at $ \sqrt{\textrm{s}} $ = 13 TeV}",
    eprint = "2403.10341",
    archivePrefix = "arXiv",
    primaryClass = "hep-ex",
    reportNumber = "CMS-HIG-18-026, CERN-EP-2024-028",
    doi = "10.1007/JHEP06(2024)097",
    journal = "JHEP",
    volume = "06",
    pages = "097",
    year = "2024"
}

@article{Cao:2021ljw,
  author        = {Cao, Junjie and Li, Demin and Lian, Jingwei and Yue, Yuanfang and Zhou, Haijing},
  title         = {{Singlino-dominated dark matter in general NMSSM}},
  eprint        = {2102.05317},
  archiveprefix = {arXiv},
  primaryclass  = {hep-ph},
  month         = {2},
  year          = {2021}
}

@article{Gunion:1984yn,
  author       = {Gunion, J. F. and Haber, Howard E.},
  title        = {{Higgs Bosons in Supersymmetric Models. 1.}},
  reportnumber = {SLAC-PUB-3404},
  doi          = {10.1016/0550-3213(86)90340-8},
  journal      = {Nucl. Phys. B},
  volume       = {272},
  pages        = {1},
  year         = {1986},
  note         = {[Erratum: Nucl.Phys.B 402, 567--569 (1993)]}
}

@article{Haber:1984rc,
  author       = {Haber, Howard E. and Kane, Gordon L.},
  title        = {{The Search for Supersymmetry: Probing Physics Beyond the Standard Model}},
  reportnumber = {UM-HE-TH-83-17, SCIPP-85-47},
  doi          = {10.1016/0370-1573(85)90051-1},
  journal      = {Phys. Rept.},
  volume       = {117},
  pages        = {75--263},
  year         = {1985}
}

@article{Khosa:2020zar,
  author        = {Khosa, Charanjit K. and Kraml, Sabine and Lessa, Andre and Neuhuber, Philipp and Waltenberger, Wolfgang},
  title         = {{SModelS database update v1.2.3}},
  eprint        = {2005.00555},
  archiveprefix = {arXiv},
  primaryclass  = {hep-ph},
  doi           = {10.31526/lhep.2020.158},
  month         = {5},
  year          = {2020}
}

@article{Martin:1997ns,
  author        = {Martin, Stephen P.},
  editor        = {Kane, Gordon L.},
  title         = {{A Supersymmetry primer}},
  eprint        = {hep-ph/9709356},
  archiveprefix = {arXiv},
  reportnumber  = {FERMILAB-PUB-97-425-T},
  doi           = {10.1142/9789812839657_0001},
  journal       = {Adv. Ser. Direct. High Energy Phys.},
  volume        = {18},
  pages         = {1--98},
  year          = {1998}
}

@article{Miller:2003ay,
  author        = {Miller, D. J. and Nevzorov, R. and Zerwas, P. M.},
  title         = {{The Higgs sector of the next-to-minimal supersymmetric standard model}},
  eprint        = {hep-ph/0304049},
  archiveprefix = {arXiv},
  reportnumber  = {CERN-TH-2003-077, DESY-03-066, ITEP-5-03},
  doi           = {10.1016/j.nuclphysb.2003.12.021},
  journal       = {Nucl. Phys. B},
  volume        = {681},
  pages         = {3--30},
  year          = {2004}
}

@article{Nilles:1983ge,
  author       = {Nilles, Hans Peter},
  title        = {{Supersymmetry, Supergravity and Particle Physics}},
  reportnumber = {UGVA-DPT-1983-12-412},
  doi          = {10.1016/0370-1573(84)90008-5},
  journal      = {Phys. Rept.},
  volume       = {110},
  pages        = {1--162},
  year         = {1984}
}

@article{vanBeekveld:2019tqp,
  author        = {van Beekveld, Melissa and Caron, Sascha and Ruiz de Austri, Roberto},
  title         = {{The current status of fine-tuning in supersymmetry}},
  eprint        = {1906.10706},
  archiveprefix = {arXiv},
  primaryclass  = {hep-ph},
  reportnumber  = {Nikhef/2019-023},
  doi           = {10.1007/JHEP01(2020)147},
  journal       = {JHEP},
  volume        = {01},
  pages         = {147},
  year          = {2020}
}

@article{Maniatis:2009re,
    author = "Maniatis, M.",
    title = "{The Next-to-Minimal Supersymmetric extension of the Standard Model reviewed}",
    eprint = "0906.0777",
    archivePrefix = "arXiv",
    primaryClass = "hep-ph",
    reportNumber = "HD-THEP-09-9",
    doi = "10.1142/S0217751X10049827",
    journal = "Int. J. Mod. Phys. A",
    volume = "25",
    pages = "3505--3602",
    year = "2010"
}

@article{Cao:2022ovk,
    author = "Cao, Junjie and Jia, Xinglong and Meng, Lei and Yue, Yuanfang and Zhang, Di",
    title = "{Status of the singlino-dominated dark matter in general Next-to-Minimal Supersymmetric Standard Model}",
    eprint = "2210.08769",
    archivePrefix = "arXiv",
    primaryClass = "hep-ph",
    doi = "10.1007/JHEP03(2023)198",
    journal = "JHEP",
    volume = "03",
    pages = "198",
    year = "2023"
}

@article{CMS2024,
    author = "Hayrapetyan, Aram and others",
    collaboration = "CMS",
    title = "{Search for exotic decays of the Higgs boson to a pair of pseudoscalars in the $\mu\mu$bb and $\tau\tau$bb final states}",
    eprint = "2402.13358",
    archivePrefix = "arXiv",
    primaryClass = "hep-ex",
    reportNumber = "CMS-HIG-22-007, CERN-EP-2023-284",
    doi = "10.1140/epjc/s10052-024-12727-4",
    journal = "Eur. Phys. J. C",
    volume = "84",
    number = "5",
    pages = "493",
    year = "2024"
}

@article{ATLAS2024,
    author = "Aad, Georges and others",
    collaboration = "ATLAS",
    title = "{Search for decays of the Higgs boson into a pair of pseudoscalar particles decaying into bb\textasciimacron{}\ensuremath{\tau}+\ensuremath{\tau}- using pp collisions at s=13\,\,TeV with the ATLAS detector}",
    eprint = "2407.01335",
    archivePrefix = "arXiv",
    primaryClass = "hep-ex",
    reportNumber = "CERN-EP-2024-164",
    doi = "10.1103/PhysRevD.110.052013",
    journal = "Phys. Rev. D",
    volume = "110",
    number = "5",
    pages = "052013",
    year = "2024"
}

@article{95GeV2023,
    author = "Cao, Junjie and Jia, Xinglong and Lian, Jingwei and Meng, Lei",
    title = "{95~GeV diphoton and $b \bar b$ excesses in the general next-to-minimal supersymmetric standard model}",
    eprint = "2310.08436",
    archivePrefix = "arXiv",
    primaryClass = "hep-ph",
    doi = "10.1103/PhysRevD.109.075001",
    journal = "Phys. Rev. D",
    volume = "109",
    number = "7",
    pages = "075001",
    year = "2024"
}

@article{Cao:2022chy,
    author = "Cao, Junjie and Lian, Jingwei and Pan, Yusi and Yue, Yuanfang and Zhang, Di",
    title = "{Impact of recent (g \ensuremath{-} 2)$_{\mu}$ measurement on the light CP-even Higgs scenario in general Next-to-Minimal Supersymmetric Standard Model}",
    eprint = "2201.11490",
    archivePrefix = "arXiv",
    primaryClass = "hep-ph",
    doi = "10.1007/JHEP03(2022)203",
    journal = "JHEP",
    volume = "03",
    pages = "203",
    year = "2022"
}

@article{CMS:2017dmg,
    author = "Khachatryan, V. and others",
    collaboration = "CMS",
    title = "{Search for light bosons in decays of the 125 GeV Higgs boson in proton-proton collisions at $ \sqrt{s}=8 $ TeV}",
    eprint = "1701.02032",
    archivePrefix = "arXiv",
    primaryClass = "hep-ex",
    reportNumber = "CMS-HIG-16-015, CERN-EP-2016-292",
    doi = "10.1007/JHEP10(2017)076",
    journal = "JHEP",
    volume = "10",
    pages = "076",
    year = "2017"
}

@article{ATLAS:2018emt,
    author = "Aaboud, Morad and others",
    collaboration = "ATLAS",
    title = "{Search for Higgs boson decays into a pair of light bosons in the $bb\mu\mu$ final state in $pp$ collision at $\sqrt{s} = $13 TeV with the ATLAS detector}",
    eprint = "1807.00539",
    archivePrefix = "arXiv",
    primaryClass = "hep-ex",
    reportNumber = "CERN-EP-2018-153",
    doi = "10.1016/j.physletb.2018.10.073",
    journal = "Phys. Lett. B",
    volume = "790",
    pages = "1--21",
    year = "2019"
}

@article{CMS:2018zvv,
    author = "Sirunyan, Albert M and others",
    collaboration = "CMS",
    title = "{Search for an exotic decay of the Higgs boson to a pair of light pseudoscalars in the final state with two b quarks and two $\tau$ leptons in proton-proton collisions at $\sqrt{s}=$ 13 TeV}",
    eprint = "1805.10191",
    archivePrefix = "arXiv",
    primaryClass = "hep-ex",
    reportNumber = "CMS-HIG-17-024, CERN-EP-2018-089",
    doi = "10.1016/j.physletb.2018.08.057",
    journal = "Phys. Lett. B",
    volume = "785",
    pages = "462",
    year = "2018"
}

@article{ATLAS:2024nnm,
    author = "Aad, Georges and others",
    collaboration = "ATLAS",
    title = "{Search for Higgs boson decays into a pair of pseudoscalar particles in the $\gamma\gamma\tau_{\text{had}}\tau_{\text{had}}$ final state using $pp$ collisions at $\sqrt{s}=13$ TeV with the ATLAS detector}",
    eprint = "2412.14046",
    archivePrefix = "arXiv",
    primaryClass = "hep-ex",
    reportNumber = "CERN-EP-2024-323",
    month = "12",
    year = "2024"
}

@article{ATLAS:2015hpr,
    author = "Aad, Georges and others",
    collaboration = "ATLAS",
    title = "{Search for new light gauge bosons in Higgs boson decays to four-lepton final states in $pp$ collisions at $\sqrt{s}=8$ TeV with the ATLAS detector at the LHC}",
    eprint = "1505.07645",
    archivePrefix = "arXiv",
    primaryClass = "hep-ex",
    reportNumber = "CERN-PH-EP-2015-111",
    doi = "10.1103/PhysRevD.92.092001",
    journal = "Phys. Rev. D",
    volume = "92",
    number = "9",
    pages = "092001",
    year = "2015"
}

@article{CMS:2012qms,
    author = "Chatrchyan, Serguei and others",
    collaboration = "CMS",
    title = "{Search for a Non-Standard-Model Higgs Boson Decaying to a Pair of New Light Bosons in Four-Muon Final States}",
    eprint = "1210.7619",
    archivePrefix = "arXiv",
    primaryClass = "hep-ex",
    reportNumber = "CMS-EXO-12-012, CERN-PH-EP-2012-292",
    doi = "10.1016/j.physletb.2013.09.009",
    journal = "Phys. Lett. B",
    volume = "726",
    pages = "564--586",
    year = "2013"
}

@article{CMS:2015nay,
    author = "Khachatryan, V. and others",
    collaboration = "CMS",
    title = "{A search for pair production of new light bosons decaying into muons}",
    eprint = "1506.00424",
    archivePrefix = "arXiv",
    primaryClass = "hep-ex",
    reportNumber = "CMS-HIG-13-010, CERN-PH-EP-2015-116",
    doi = "10.1016/j.physletb.2015.10.067",
    journal = "Phys. Lett. B",
    volume = "752",
    pages = "146--168",
    year = "2016"
}

@article{ATLAS:2018coo,
    author = "Aaboud, Morad and others",
    collaboration = "ATLAS",
    title = "{Search for Higgs boson decays to beyond-the-Standard-Model light bosons in four-lepton events with the ATLAS detector at $\sqrt{s}=13$ TeV}",
    eprint = "1802.03388",
    archivePrefix = "arXiv",
    primaryClass = "hep-ex",
    reportNumber = "CERN-EP-2017-293",
    doi = "10.1007/JHEP06(2018)166",
    journal = "JHEP",
    volume = "06",
    pages = "166",
    year = "2018"
}

@article{CMS:2018jid,
    author = "Sirunyan, Albert M and others",
    collaboration = "CMS",
    title = "{A search for pair production of new light bosons decaying into muons in proton-proton collisions at 13 TeV}",
    eprint = "1812.00380",
    archivePrefix = "arXiv",
    primaryClass = "hep-ex",
    reportNumber = "CMS-HIG-18-003, CERN-EP-2018-288",
    doi = "10.1016/j.physletb.2019.07.013",
    journal = "Phys. Lett. B",
    volume = "796",
    pages = "131--154",
    year = "2019"
}

@article{Chatterjee:2022pxf,
    author = "Chatterjee, Arindam and Datta, AseshKrishna and Roy, Subhojit",
    title = "{Electroweak phase transition in the Z$_{3}$-invariant NMSSM: Implications of LHC and Dark matter searches and prospects of detecting the gravitational waves}",
    eprint = "2202.12476",
    archivePrefix = "arXiv",
    primaryClass = "hep-ph",
    reportNumber = "HRI-RECAPP-2022-001",
    doi = "10.1007/JHEP06(2022)108",
    journal = "JHEP",
    volume = "06",
    pages = "108",
    year = "2022"
}

@article{Choi:2019yrv,
    author = "Choi, Kiwoon and Im, Sang Hui and Jeong, Kwang Sik and Park, Chan Beom",
    title = "{Light Higgs bosons in the general NMSSM}",
    eprint = "1906.03389",
    archivePrefix = "arXiv",
    primaryClass = "hep-ph",
    reportNumber = "CTPU-PTC-19-17, PNUTP-19-A11",
    doi = "10.1140/epjc/s10052-019-7473-1",
    journal = "Eur. Phys. J. C",
    volume = "79",
    number = "11",
    pages = "956",
    year = "2019"
}

@article{LI2024116591,
title = {Impact of the LZ experiment on the DM phenomenology and naturalness in the MSSM},
journal = {Nuclear Physics B},
volume = {1005},
pages = {116591},
year = {2024},
issn = {0550-3213},
doi = {https://doi.org/10.1016/j.nuclphysb.2024.116591},
url = {https://www.sciencedirect.com/science/article/pii/S0550321324001573},
author = {Dongwei Li and Lei Meng and Haijing Zhou}
}

@article{Ellwanger:2011mu,
    author = "Ellwanger, Ulrich and Espitalier-Noel, Gregory and Hugonie, Cyril",
    title = "{Naturalness and Fine Tuning in the NMSSM: Implications of Early LHC Results}",
    eprint = "1107.2472",
    archivePrefix = "arXiv",
    primaryClass = "hep-ph",
    reportNumber = "LPT-ORSAY-11-63, L2C-11, LUPM-11-002, L2C-11-",
    doi = "10.1007/JHEP09(2011)105",
    journal = "JHEP",
    volume = "09",
    pages = "105",
    year = "2011"
}

@article{Fowlie:2024nhs,
    author = "Fowlie, Andrew and Herrera, Gonzalo",
    title = "{Precise interpretations of traditional fine-tuning measures}",
    eprint = "2406.03533",
    archivePrefix = "arXiv",
    primaryClass = "hep-ph",
    doi = "10.1103/PhysRevD.111.015020",
    journal = "Phys. Rev. D",
    volume = "111",
    number = "1",
    pages = "015020",
    year = "2025"
}

@article{Baer:2026idw,
    author = "Baer, Howard and Barger, Vernon and Sengupta, Dibyashree",
    title = "{Can blind spots save neutralino dark matter in natural supersymmetry models?}",
    eprint = "2607.05648",
    archivePrefix = "arXiv",
    primaryClass = "hep-ph",
    reportNumber = "OU-HEP-260702",
    month = "7",
    year = "2026"
}

@article{Cepeda:2019klc,
    author = "Cepeda, M. and others",
    editor = "Dainese, Andrea and Mangano, Michelangelo and Meyer, Andreas B. and Nisati, Aleandro and Salam, Gavin and Vesterinen, Mika Anton",
    title = "{Report from Working Group 2}: {Higgs Physics at the HL-LHC and HE-LHC}",
    eprint = "1902.00134",
    archivePrefix = "arXiv",
    primaryClass = "hep-ph",
    reportNumber = "CERN-LPCC-2018-04",
    doi = "10.23731/CYRM-2019-007.221",
    journal = "CERN Yellow Rep. Monogr.",
    volume = "7",
    pages = "221--584",
    year = "2019"
}

@article{ATLAS:2025qyn,
    author = "Aad, Georges and others",
    collaboration = "ATLAS",
    title = "{Search for Higgs boson exotic decays into Lorentz-boosted light bosons in the four-{\ensuremath{\tau}} final state at s=13TeV with the ATLAS detector}",
    eprint = "2503.05463",
    archivePrefix = "arXiv",
    primaryClass = "hep-ex",
    reportNumber = "CERN-EP-2025-038",
    doi = "10.1016/j.physletb.2025.139843",
    journal = "Phys. Lett. B",
    volume = "870",
    pages = "139843",
    year = "2025"
}

@article{CMS:2025hjt,
    author = "Chekhovsky, Vladimir and others",
    collaboration = "CMS",
    title = "{Search for light pseudoscalar boson pairs produced from Higgs boson decays using the 4{\ensuremath{\tau}} and 2{\ensuremath{\mu}}2{\ensuremath{\tau}} final states in proton-proton collisions at $ \sqrt{s}=13 $ TeV}",
    eprint = "2508.06947",
    archivePrefix = "arXiv",
    primaryClass = "hep-ex",
    reportNumber = "CMS-SUS-24-002, CERN-EP-2025-129",
    doi = "10.1007/JHEP04(2026)132",
    journal = "JHEP",
    volume = "04",
    pages = "132",
    year = "2026"
}

@article{Hambye:2007vf,
    author = "Hambye, Thomas and Tytgat, Michel H. G.",
    title = "{Electroweak symmetry breaking induced by dark matter}",
    eprint = "0707.0633",
    archivePrefix = "arXiv",
    primaryClass = "hep-ph",
    reportNumber = "ULB-TH-07-26",
    doi = "10.1016/j.physletb.2007.11.069",
    journal = "Phys. Lett. B",
    volume = "659",
    pages = "651--655",
    year = "2008"
}

@article{Dobrescu:2026fok,
    author = "Dobrescu, Bogdan A. and Roy, Subhojit",
    title = "{Axion-like particle at the 10 GeV scale: Higgs decays to wide jets and photons}",
    eprint = "2608.18506",
    archivePrefix = "arXiv",
    primaryClass = "hep-ph",
    reportNumber = "FERMILAB-PUB-26-0620-T",
    month = "8",
    year = "2026"
}

@article{Abdallah:2020yag,
    author = "Abdallah, Waleed and Datta, AseshKrishna and Roy, Subhojit",
    title = "{A relatively light, highly bino-like dark matter in the Z$_{3}$-symmetric NMSSM and recent LHC searches}",
    eprint = "2012.04026",
    archivePrefix = "arXiv",
    primaryClass = "hep-ph",
    reportNumber = "HRI-RECAPP-2020-10",
    doi = "10.1007/JHEP04(2021)122",
    journal = "JHEP",
    volume = "04",
    pages = "122",
    year = "2021"
}

@article{Roy:2024yoh,
    author = "Roy, Subhojit and Wagner, Carlos E. M.",
    title = "{Dark Matter searches with photons at the LHC}",
    eprint = "2401.08917",
    archivePrefix = "arXiv",
    primaryClass = "hep-ph",
    reportNumber = "EFI 24-1",
    doi = "10.1007/JHEP04(2024)106",
    journal = "JHEP",
    volume = "04",
    pages = "106",
    year = "2024"
}

@article{Arganda:2025fhx,
    author = "Arganda, Ernesto and de los Rios, Mart{\'\i}n and Perez, Andres D. and Roy, Subhojit and Sand{\'a} Seoane, Rosa M. and Wagner, Carlos E. M.",
    title = "{Shedding light on dark matter at the LHC with machine learning}",
    eprint = "2509.15121",
    archivePrefix = "arXiv",
    primaryClass = "hep-ph",
    reportNumber = "EFI-25-12, IFT-UAM/CSIC-25-93",
    doi = "10.1103/m1cd-1sfb",
    journal = "Phys. Rev. D",
    volume = "113",
    number = "9",
    pages = "095013",
    year = "2026"
}

@article{Datta:2022bvg,
    author = "Datta, AseshKrishna and Guchait, Monoranjan and Roy, Arnab and Roy, Subhojit",
    title = "{Hunting ewinos and a light scalar of Z$_{3}$-NMSSM with a bino-like dark matter in top squark decays at the LHC}",
    eprint = "2211.05905",
    archivePrefix = "arXiv",
    primaryClass = "hep-ph",
    reportNumber = "HRI-RECAPP-2022-014",
    doi = "10.1007/JHEP11(2023)081",
    journal = "JHEP",
    volume = "11",
    pages = "081",
    year = "2023"
}

\end{document}